\RequirePackage{fix-cm}
\documentclass{svjour3}       
\smartqed  
\usepackage{graphicx}
\usepackage{todonotes}
\usepackage{amsmath}
\usepackage{booktabs}
\usepackage{tabularx}
\usepackage{longtable}
\usepackage{array}
\usepackage{multirow}
\usepackage{rotating}
\usepackage{tcolorbox}
\usepackage{float}
\usepackage{listings}

\usepackage{hyperref}

\begin{document}

\title{On the Relation between Code Quality and Machine Learning Performance: A Large-scale Empirical Study}


\author{Marius Mignard \and Steven Costiou~$^*$ \and Anne Etien~$^*$.}

\def\thefootnote{*}\footnotetext{These authors are both last author}


\institute{Marius Mignard \at
              Univ. Lille, Inria, CNRS, Centrale Lille, UMR 9189 CRIStAL F-59000 Lille, France \\
              \email{marius.mignard@inria.fr}           
           \and
           Steven Costiou \at
              UMR 9189 CRIStAL, Univ. Lille, Inria, CNRS, Centrale Lille
              \email{steven.costiou@inria.fr}  
            \and
            Anne Etien \at
            UMR 9189 CRIStAL, Univ. Lille, CNRS, Inria, Centrale Lille
            \email{anne.etien@inria.fr}
}

\date{Received: date / Accepted: date}

\maketitle

\begin{abstract}
\textbf{Context:} Computational notebooks are the standard environment for machine learning (ML) development. Within the ML community, model performance is often the primary considered metric, and code quality is treated as a secondary concern. This prioritization relies on a largely untested assumption that code quality and ML performance are unrelated. Practitioners also reuse existing code that may come from notebooks selected through social signals (popularity, author expertise) whose reliability as quality proxies has never been assessed. 

\noindent
\textbf{Objective:} We empirically investigated the relationship between code quality and ML performance in notebooks, and evaluated whether popularity and author expertise give indication on code quality or performance. 

\noindent
\textbf{Method:} We conducted a large-scale empirical study of 265,363 Python notebooks submitted to Kaggle competitions. We assessed code quality with two static analysis tools: Pylint, capturing general Python code quality, and SonarQube, configured with a profile of 34 rules targeting data-science and ML-specific practices. 

\noindent
\textbf{Results:} The relationship between code quality and performance depends on the notion of quality considered. General Python code quality is decoupled from ML performance, showing negligible or non-significant correlations across all observations. In contrast, ML-specific violations exhibit a consistent, small negative association with performance that persists across all observations. The popularity of a notebook does not give information on the code quality or performance. Code expertise provides no information on quality or performance, but competition expertise correlates with better performance, fewer ML-specific violations, and slightly more Python errors and refactoring violations. 

\noindent
\textbf{Conclusions:} Following general software engineering best practices does not impact ML performance, while adhering to ML-specific practices is associated with better ML performance. Software engineering best practices should thus be integrated from the earliest stages of ML development, as the gains in reproducibility, maintainability, and comprehension will not negatively impact ML performance. Practitioners and researchers should not rely on popularity or expertise when selecting notebooks for reuse or sampling.

\end{abstract}


\section{Introduction}
\label{sec:introduction}
Computational notebooks have become the standard environment for machine learning (ML) and data science, adopted across research, industry, and education~\cite{rule18exploration,chattopadhyay2020s,quaranta2022a,zhang2020data}. 
Their cell-based, interactive design favors an iterative~\cite{rule18exploration}, "trial-and-error" workflow that prioritizes rapid experimentation~\cite{quaranta2022a}. 
However, this flexibility may come at a cost: a departure from established software engineering (SE) best practices.
Within the ML community, the primary metric of success is often the model's performance, frequently leading practitioners to treat code quality as a secondary concern~\cite{jebnoun2020scent,kim2016emerging,quaranta2022a,sculley2015hidden}. 
This prioritization implicitly relies on a prevailing, yet largely untested, assumption: that "clean code" and "high performance" are distinct and unrelated factors.
Such an assumption is risky. 
By neglecting code quality, practitioners may introduce technical debt that undermines the reliability, understandability, and reproducibility of the code~\cite{sculley2015hidden,quaranta2022a,jebnoun2020scent,dong2021splitting}. 
This could also impact the model's performance itself if "clean code" and "high performance" are, in fact, related. 

Moreover, code quality directly affects the agility of ML workflows in industrial settings~\cite{lanubile2021towards}.
The transition from an experimental notebook to production-ready code is frequently hindered by technical debt~\cite{chattopadhyay2020s,dong2021splitting,sculley2015hidden,pertseva24theory}, requiring extensive refactoring or rewriting by software engineers, and leading to increased costs and delayed deployment. 

In this paper, we investigate, in the context of notebooks, the correlation between code violations and ML performance to determine whether poor coding practices represent a hidden threat to model outcomes, or if high-quality code is compatible with state-of-the-art machine learning results. 
If achieving high performance does not necessitate a sacrifice in code quality, then software engineering best practices should be integrated from the earliest stages of experimentation rather than treated as a costly afterthought.

The implications of this question extend beyond individual notebooks. 
Code reuse is a common practice in the ML community~\cite{kallen2020jupyter,yang2023code,jebnoun2022clones,koenzen2020code,chattopadhyay2020s}.
Practitioners frequently adapt existing notebooks, following a "clone and own" culture, and platforms like Kaggle\footnote{\url{https://www.kaggle.com/}} serve as pedagogical resources~\cite{stein2024learning}. 
If performance is not a reliable proxy for code quality, practitioners who prioritize performance may unknowingly propagate low-quality code throughout the ecosystem. 
When highly-upvoted notebooks, or notebooks authored by experts, frequently violate coding standards, they risk institutionalizing poor habits among the novices who reuse them as pedagogical references. 
Understanding the relationship between expertise, popularity, and quality is therefore essential both for code reuse decisions and for data science education.

To our knowledge, the relationship between code quality and ML performance has never been empirically tested at scale, and neither has the reliability of the social signals (popularity, author expertise) that practitioners might use to select notebooks to reuse. 
This paper addresses this gap. 
We conduct a large-scale empirical study of $265,363$ Python notebooks submitted to Kaggle competitions. 
Kaggle provides a unique setting for this investigation: each notebook is associated with a competition score that serves as a measurable proxy for ML performance, together with popularity metrics (upvotes) and a user expertise system (progression tiers). 
We assess notebook quality with two complementary static analysis tools: Pylint, capturing general Python code quality, and SonarQube, configured with a profile of 34 rules targeting data-science and ML-specific practices.
In particular, we address the following research questions (RQs):
\begin{itemize}
    \item \textbf{RQ1.} Is there a correlation between notebook popularity and code quality?
    \item \textbf{RQ2.} How do violations of best practices relate to machine learning performance?
    \item \textbf{RQ3.} Do expert and non-expert users differ in their violation of best practices?
\end{itemize}
Our results show that the performance-quality dichotomy is neither simply true nor simply false.
The relation between those factors depends on which notion of quality is considered. 
General Python code quality, as captured by Pylint, is effectively decoupled from ML performance. 
In contrast, ML-specific violations exhibit a consistent negative association with performance, which persists across competition types and expertise levels. 
We further find that neither popularity nor code expertise reliably signals quality or performance, challenging the trust practitioners place in these social signals.
This paper makes the following contributions:
\begin{itemize}
    \item A large-scale empirical study of the relationship between code quality and ML performance on 265,363 Kaggle competition notebooks, combining general (Pylint) and ML-specific (SonarQube) static analysis;
    \item Empirical evidence that general Python quality is decoupled from ML performance, whereas adherence to ML-specific practices is associated with better performance outcomes;
    \item An analysis of the reliability of Kaggle's social signals (popularity, expertise tiers) as proxies for code quality and performance;
\end{itemize}

The remainder of this paper is organized as follows. Section~\ref{sec:background} provides an overview of the relevant state of the art.
Section~\ref{dataset_building} describes the construction of our dataset. 
Section~\ref{sec:study_design} presents our study design and statistical framework. 
Section~\ref{sec:results} reports the results for each research question, and Section~\ref{sec:discussion} discusses their implications. 
Section~\ref{sec:threats} examines the threats to the validity of our study, and Section~\ref{sec:conclusion} concludes.

\section{State of the Art}
\label{sec:background}

\subsection{Code Quality of Python and ML Notebooks}
Code quality is a multi-faceted notion covering maintainability, readability, and reliability. 
A common way to implement it is through 
sets of rules (e.g., PEP8~\cite{pep8}) whose violations can be detected by static analysis, without executing the code. 
The remainder of the section reviews some of the lines of work assessing code quality in Python notebooks and ML code through static analysis. All these works measure or detect the \emph{prevalence} of violations and smells; none relates them to the \emph{performance} of the ML model the code produces.

\paragraph{Notebooks versus scripts.}
Grotov et al.~\cite{grotov2022a} compared structural and stylistic metrics between Python scripts and notebooks collected on GitHub, using Hyperstyle~\cite{hyperstyle}, a tool combining the Pylint, Flake8, and WPS static analyzers. 
They found that notebooks contain more quality rule violations than scripts and advocated for notebook-specific quality tools. 
Adams et al.~\cite{adams2023a} replicated this study on Kaggle, with 12,136 scripts and 57,722 notebooks, and reached the opposite conclusion: in their corpus, scripts exhibit lower code quality.

\paragraph{Large-scale studies of notebook quality.}
Wang et al.~\cite{wang2020b} argued for the need to analyze notebook quality, supported by a preliminary study of 1,982 high-quality notebooks (example notebooks curated by the Jupyter team) combining a PEP8 checker, AST-based detection of unused variables, and deprecated API usage. 
Siddik and Bezemer~\cite{siddik2023a} linted 246,599 Kaggle notebooks with Pylint and found that non-ML notebooks display slightly better code quality than ML ones. 
Quaranta et al.~\cite{quaranta2022a} proposed a catalog of best practices for collaborative notebooks, one of which is writing high-quality code. 
They assessed code quality best practices by linting 1,380 collaboratively authored, high-quality Kaggle notebooks, in which 74.49\% violated at least one convention rule, 68.39\% raised warnings, 55.74\% raised errors, and 29.58\% had refactor violations. 
They also observed a slightly better compliance with best practices among the most upvoted notebooks of their sample. 
Van Oort et al.~\cite{van21a} also studied the prevalence of code smells in 74 Python ML projects using Pylint.

\paragraph{ML-specific code smells.}
Beyond general-purpose rules, previous work characterizes quality issues specific to ML code. 
Jebnoun et al.~\cite{jebnoun2020scent} compared the prevalence of ten traditional code smells (e.g., Long Parameter List, Large Class) between 59 deep learning and 59 traditional systems from
GitHub using PySmell~\footnote{\url{https://github.com/chenzhifei731/Pysmell}}, and found no significant difference between the two populations. Zhang et al.~\cite{zhang2022a} defined a list of ML-specific code smells (e.g., Columns and DataType Not Explicitly Set, DataFrame Conversion API Misused, In-Place APIs Misused). 
Building on this list, Recupito et al.~\cite{recupito2025a} introduced CodeSmile, a detector covering 12 of these smells, and analyzed 337 ML projects from the NICHE dataset~\cite{widyasari2023a}.
Recupito et al. observed that ML smells are frequently introduced when files are modified for new features, are typically removed during feature
enhancement, or refactoring tasks, and are mostly resolved within the first 10\% of commits. 
In an educational context, Skripchuk et al.~\cite{skripchuk2022a} analyzed 19 ML student projects and, among other violations, found that all contained Python code smells, while 14 lacked systematic hyperparameter tuning, 7 used test data outside of model evaluation, and 6 confused classification with regression.

\paragraph{Notebook-aware and ML-aware quality tools.}
These studies converge on one observation: general-purpose linters cannot detect bad practices specific to ML workflows, such as non-reproducible random number generation, silent type inference on data loading, or API misuses that alter the behaviour of a pipeline~\cite{van21a,wang2020b}. 
Dedicated tools have emerged on both notebook specific and ML specific sides. 
On the notebook side, Julynter~\cite{pimentel2021a} checks 21 rules targeting notebook-specific issues (hidden state, imports, cell titles), Pynblint~\cite{quaranta2024a} verifies the best practices
identified by Quaranta et al.~\cite{quaranta2022a}, and NBLizer~\cite{subotic2022a} performs static analysis aware of out-of-order cell execution, with a data leakage check evaluated on 2,211
Kaggle notebooks. 
On the ML side, dslinter~\cite{haakman2020a} extends Pylint with ML-specific rules, mllint~\cite{van2022a} assesses the software quality of ML projects, and MLScent~\cite{shivashankar2025a}
implements 76 ML-specific rules. 
Pyra~\cite{dolcetti2026a} takes a different route: it reasons over a domain-specific type system built on abstract interpretation to check 16 rules (e.g., data leakage, reproducibility, plotting misuses), and was evaluated on 2,286 Kaggle notebooks, yielding 4,214 violations. 
Data-centric tools such as Pandera\footnote{\url{https://pandera.readthedocs.io/}} and the Data
Linter~\cite{hynes2017a} focus on validating the data rather than the code.
In parallel, SonarQube, a leading industrial static analysis platform~\cite{marc19a}, introduced a set of rules dedicated to data science libraries (pandas, NumPy, scikit-learn, TensorFlow, PyTorch).\newline

\noindent
Prior work has characterized the prevalence of 
quality issues in notebooks and ML code, and has produced specialized detection tools. 
However, whether code quality relates to 
the outcomes of a notebook, such as the performance of the resulting model remains an open question, 
which no study has yet investigated at scale. 
Our study addresses this gap on 265,363 Kaggle notebooks, using Pylint for 
general Python quality and SonarQube for ML-specific quality.

\subsection{Social Signals and Kaggle as a Study Platform}
\label{sec:bg-social}

Practitioners may rely on social signals (votes, reputation, expertise\dots) to decide which code to trust and reuse. 
Evidence from Stack Overflow\footnote{\url{https://stackoverflow.com/questions}} suggests this trust is fragile: Zhang et al.~\cite{zhang2018a} found
that even highly upvoted answers frequently contain API misuses; Bafatakis et al.~\cite{bafatakis2019a} found that Python code snippets containing fewer code violations tend to be more upvoted than the ones with more violations but found no correlation between users' reputation and compliance with coding best practices.
Similarly, Rahman et al.~\cite{rahman2019a} found no correlation between the users' reputation and the introduction of insecure Python snippets in Stack Overflow answers.

Fischer et al.~\cite{fischer2017a} showed that insecure snippets from popular answers propagate into production applications. 
On Kaggle, the evidence is scarcer. 
Quaranta et al.~\cite{quaranta2022a} observed a slight tendency of the most upvoted notebooks to better comply with best practices, on a small, pre-filtered sample. 
Mostafavi Ghahfarokhi et al.~\cite{ghahfarokhi2025a} leveraged community feedback on Kaggle notebooks to predict code understandability from static metrics, treating user opinions as a quality oracle rather than testing their reliability. 
Whether upvotes and expertise tiers actually signal quality at the scale of the platform remains an open question.\newline


\noindent
Several datasets support large-scale notebook research. 
KGTorrent~\cite{quaranta2021a} first assembled a corpus of Kaggle notebooks with their metadata; Kaggle now publishes this information through Meta Kaggle and Meta Kaggle Code~\cite{megan_risdal_timo_bozsolik_2022,jim_plotts_megan_risdal_2023}, on which our study is built.
Code4ML~\cite{drozdova2023a} annotates Kaggle code snippets with ML
pipeline stages. To our knowledge, no existing dataset combines static
analysis violations, competition performance scores, and social metadata
at the notebook level.

\section{Notebook Selection Methodology and Dataset Building}
\label{dataset_building}
This section describes our process to select the notebooks, extract the violations, and compute metrics.
All the implementations, configurations, and versions of the tools used can be found in the reproduction package \url{https://doi.org/10.5281/zenodo.21464700}.
The compute-intensive tasks have been run on Grid5000 infrastructure~\cite{grid5000} on the Chirop cluster (DL360 Gen10+, Xeon Platinum 8358, 512 GiB of RAM, and Debian 5.10).

\subsection{Data and Tool Choice}\label{sec:tool_choice}


\paragraph{Kaggle as our data source}
We selected Kaggle as our data source for several reasons. 
As one of the largest platforms for Machine Learning, Kaggle provides an ecosystem where notebooks (written in R or Python) are linked to data science tasks. 
Unlike GitHub, where notebooks can serve diverse purposes (e.g., documentation, experimentation, personal scripts), Kaggle notebooks are domain-specific. 
This allowed us to bypass the complex filtering required to identify relevant ML notebooks on general-purpose repositories.

As we studied coding practices along with ML performance, a key requirement for our study is a metric for ML performance. 
Kaggle hosts ML competitions where users submit notebooks, each associated with a competition score that measures the performance of a given ML pipeline on a specific task (e.g., classification, prediction, or generation). 
We use this score as a proxy for ML performance. 
During a competition, participants receive feedback via a public score calculated on a subset of the test data. 
However, to discourage overfitting and ensure the model's ability to generalize, the final ranking is determined by a private score, computed on a hidden dataset revealed only after the competition ends. 
We use the private score in our analysis, as it provides a more accurate and trustworthy measure of the submitted solution's real performance than the public score.

\paragraph{Code Quality Tools.}
\label{sec:code_quality_tool_choice}

To detect violations of best practices, we based our analysis on two external tools:
\begin{itemize}
\item \textbf{Pylint\footnote{version~4.02} :} Selected for its wide acceptance across the Python community~\cite{dasg17a} and strict adherence to Python's PEP8. It identifies general coding flaws, convention issues, and Python errors. 
\item \textbf{SonarQube\footnote{sonarqube:25.11.0.114957-community \& sonar-scanner-8.0.1.6346}:} As a leading industry standard for static analysis~\cite{marc19a}, SonarQube provides a set of rules targeting Data Science libraries. This allows us to detect specific bad practices in ML workflows that general-purpose linters might miss.
\end{itemize}

\subsection{Notebook Selection}

\begin{figure}[t]
\centering
\begin{tikzpicture}[
  node distance=8mm and 14mm,
  every node/.style={
    rectangle,
    rounded corners,
    draw,
    align=center,
    font=\small,
    text width=6.8cm
  },
  dataset/.style={
    fill=gray!10, 
    very thick,  
    rounded corners=2pt 
  },
  arrow/.style={->, thick}
]

\node (start) [dataset] {Kaggle\
\textit{(snapshots from Jan. 2026)}\\
Starting point: 8M notebooks files};

\node (filterCompetiton) [below=of start] {\textbf{Filter 1. Competition} (section~\ref{filter_1_comp})\\
Notebooks submitted to competition (exclusion of \textbf{Getting Started} competitions)\\
Remaining: 355,760 notebooks};

\node (filterPython) [below=of filterCompetiton] {\textbf{Filter 2. Python} (section~\ref{filter_2_python})\\ 
Removal of scripts and \textbf{not Python notebooks}\\
Removal of \textbf{notebooks using Python versions lower than 3.6}\\
Removal of the notebooks with \textbf{less than 30 lines of Python code}\\
Remaining: 309,291 notebooks};

\node (filterTools) [below=of filterPython] {\textbf{Filter 3. Tools} (section~\ref{filter_3_tools})\\
Removal of the notebooks \textbf{failed to run on JupyText, Pylint or Sonar}\\
Remaining:  297,417 notebooks};

\node (z_score) [below=of filterTools] {\textbf{Computed Metrics} (section~\ref{z-score})\\
Removal of the notebooks where the \textbf{z-score can not be computed}\\
Remaining:  265,363 notebooks};

\node (final) [below=of z_score, dataset] {\textit{Final dataset}\\
265,363 notebooks};

\draw[arrow] (start) -- (filterCompetiton);
\draw[arrow] (filterCompetiton) -- (filterPython);
\draw[arrow] (filterPython) -- (filterTools);
\draw[arrow] (filterTools) -- (z_score);
\draw[arrow] (z_score) -- (final);

\end{tikzpicture}
\caption{Data filtering and selection process for Kaggle notebooks.}
\label{fig:selection_filters}
\end{figure}
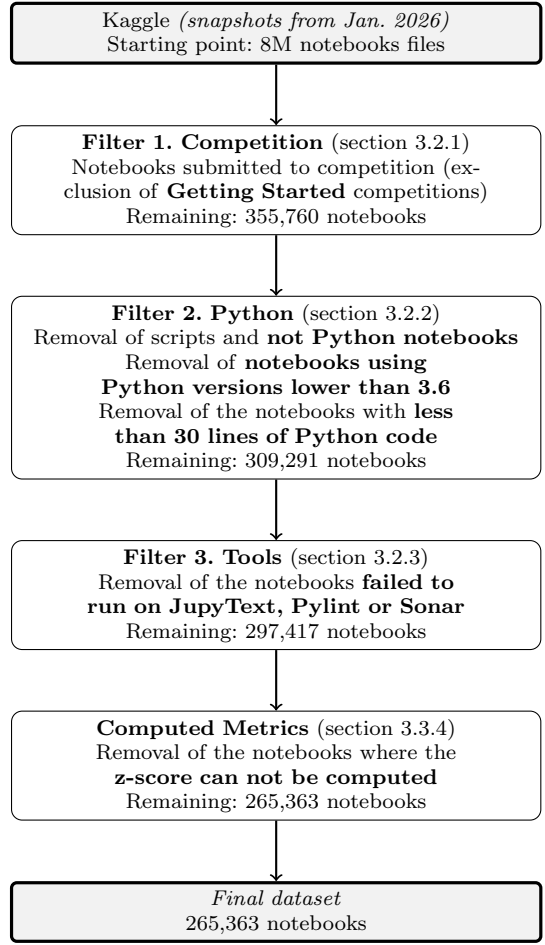

To collect the data used in this study, we relied on the Kaggle open users' data~\cite{megan_risdal_timo_bozsolik_2022,jim_plotts_megan_risdal_2023}, which contains the notebooks source files (e.g., .py, .r, and .ipynb) with their metadata.
Kaggle opened its users' public data to allow anyone to use it as study material, without the need to scrape the website.
Since the Kaggle data are updated daily, we used a snapshot that we downloaded on January 30, 2026. 
At the time of our snapshot, we had 8 million notebooks. 
We then applied three filters successively to select the final notebooks to be used in the study (Fig~\ref{fig:selection_filters}).

\subsubsection{Filter 1: Competition}
\label{filter_1_comp}

\paragraph{Notebooks submitted in competition.} Our objective is to analyze the notebooks' quality from a software engineering point of view, together with their ML-perfor\-man\-ce. 
As said before, we chose to use the Kaggle competition score as a proxy for the ML-performance. 
Thus, we needed notebooks submitted in a Kaggle competition and selected all notebooks that we were able to associate with a competition submission. 
This initial filtering yielded 414,129 notebooks, each linked to a competition submission (not all notebooks on Kaggle are authored for competition; most of them aren't). 


\paragraph{Exclusion of Getting Started competition.} Kaggle competitions belong to several segments (Playground, Featured, Recruitment, Research, Community, Getting Started, Analytics...).
We excluded Getting Started competitions because they frequently contain large numbers of nearly identical notebooks derived from tutorials, which would bias the analysis.
After these exclusions, we retained 355,760 notebooks. 

\subsubsection{Filter 2: Python}
\label{filter_2_python}

\paragraph{Exclusion of non-Python notebooks.} The notebooks on Kaggle can be authored as notebooks in Python or in R and as Python scripts. 
In this study, we chose to focus on Python code violations because it is the most widely used language for machine learning and data science \cite{brai18a} and the most used language on Kaggle (98\% of the code is written in Python).
We also do not consider code written as scripts, as it only represents a small part of the Kaggle code.
Scripts are users' notebooks converted to utility scripts for easier reuse, and they represent 4\% of the total code volume. 
We therefore excluded all the scripts and notebooks not written in Python, giving us 
325,823 notebooks. 

\paragraph{Removal of notebook with Python version older than 3.6.} Since we considered all the available notebooks, some of the oldest we found were written in Python 2. 
Each release of Python, particularly from Python 2 to 3, introduced new practices and modified language usage.
Static analysers need to rely on a given selection of versions to properly assess rules applicable to the selected versions.
We chose to retain the notebook with a Python version equal to or greater than 3.6.
Python 3.6 introduced new features of Python that have become standard (f-strings, underscored numbers, type hints\dots), and the majority of the previously selected notebooks were written in Python 3.6 or greater. 
We parsed the Python version declared in the notebooks file, and we found 643 notebooks with a version lower than 3.6 that we dropped, and 894 without a specified version.
The ones without a specified version were also dropped, leaving 324,286 notebooks.

\paragraph{Exclusion of Notebooks with less than 30 lines of code.} Notebooks can contain none or a small number of lines of code.  
Some notebooks contain only a few lines, either to generate random output (sometimes leading to great competition scores) or to upload pre-calculated results files to the competition.
The last point is valid regarding to the competition rules, but this type of code is not interesting regarding to our study since it does not contain ML pipeline code.
The competition score associated with those notebooks does not come from the notebook itself, but from inaccessible code. We considered that these notebooks would introduce bias to the study, as the code and the score are not linked.

The information on the notebook number of lines wasn't in Kaggle's metadata. 
To add the number of lines in each notebook, we used \textit{Radon}\footnote{Radon is a Python tool that computes various metrics from the source code. \url{https://radon.readthedocs.io/en/master/}} and added the resulting metrics (LOC, LLOC, SLOC, Comments). 
\textit{Radon} failed to run on $1,194$ notebooks which were subsequently dropped.


We removed all notebooks with fewer than 30 source lines of code (SLOC), which excludes blank lines, comments, and Markdown text to only consider Python code. 
While this threshold is arbitrary, it extends the methodology of~\cite{dolcetti2026pyra}, who used a 20 line limit for Kaggle notebooks. 
We introduced an additional 10 line margin to ensure the remaining notebooks contained a more substantial amount of code. 
At this step, we had 309,291 notebooks remaining. 



\subsubsection{Filter 3: Tools}
\label{filter_3_tools}

\paragraph{Notebooks conversion.}
\label{para:nb_convertion}
To use Pylint and SonarQube on the selected notebooks, we had to convert them into regular Python scripts.
We used \textit{JupyText}\footnote{\url{https://jupytext.readthedocs.io/en/latest/}} for the conversion, allowing us to escape the notebook ‘magic’ commands\footnote{Jupyter notebook ‘magic’ commands can be used for bash commands or to control the underlying Python kernel running the notebook's code (using the debugger, profile code and time execution\dots).} (lines starting with !, \% or \%\%), while keeping all the markdown contents as Python comments.
All the notebooks were successfully converted without error or warning messages.
At this point, and for the remainder of our work, we have no information on whether the code of the Python script can run or not. Unlike studies that target the reproducibility of notebooks~\cite{pimentel2019a,wang2020a}, we cannot execute the scripts because we do not reconstruct the execution environment (libraries and imported data), which would rapidly lead to errors (e.g., ImportError, FileNotFoundError\dots).

\paragraph{Pylint violations.}
\label{para:nb_convertion:pylint_violation}
On the converted notebooks, we ran Pylint file by file. 
In terms of configuration, we set the minimum Python version to be used for version-dependent checks to Python 3.6.0.
Additionally, since the third-party libraries imported by the notebooks were not installed in our evaluation environment, we configured Pylint to ignore these external dependencies. 
This prevented Pylint from triggering false positives, such as \texttt{import-error} or \texttt{no-member}, for libraries outside the scope of our analysis.


We then collected all the outputs, we counted the occurrences per rule, and added them to the dataset.
On 10 of the selected notebooks, Pylint had a fatal error stopping their analysis. 
We dropped the notebooks that we were unable to analyse. 
Pylint flagged a syntax error in 11,864 notebooks.
Some of the notebooks with syntax error still contained non Python code (e.g., \texttt{pip install cmaes}, which was incorporated into the script during the conversion step because of the missing \texttt{!} indicator). Additionally, we found indentation issues in the Python code. This faulty code come from cells that do not run; however, in a notebook environment, a single cell can output an error without blocking the execution of the remainder of the notebook. This behavior does not exist in standard Python scripts.
When a syntax error is raised by Pylint, other rules are not checked, making it the only violation of the notebook\footnote{Pylint relies on the Python AST to check rules. The AST cannot be built when there is a syntax error, meaning that other rules cannot be checked \url{https://stackoverflow.com/a/78419051}}. 
We removed the 11,864 notebooks with syntax errors, leaving 297,417 notebooks. 

\paragraph{SonarQube.} We followed a similar process for SonarQube. On the converted files, we ran the sonar scanner.
We then used a script to retrieve all the violations and add them to their corresponding notebooks in the dataset. 
We did not encounter any fatal error; the SonarQube scanner successfully scanned all the files. 

\subsection{Dataset Enrichment} 
The notebook selection steps yielded 297,417 notebooks associated with competition score, user information, and other metadata. 
We added to this dataset the following new metrics that we will use later in our analysis.

\subsubsection{Users' expertise level} 
\label{users_expertise}
Kaggle uses a progression system with five tiers: \textbf{Novice}, \textbf{Contributor}\footnote{Contributor is a legacy tier that is not used anymore by Kaggle. Users were able to get the contributor tier with regular use of the platform (i.e., running a notebook, making competition submissions, and leaving comments on other users' notebooks\ldots).}, \textbf{Expert}, \textbf{Master}, \textbf{Grand Master}. 
Each user can be one of those on the three tracks: \textbf{Code}, \textbf{Competition}, \textbf{Dataset} (a user can be a master in Code and an expert in Competition). 
To unlock each tier, depending on the track, users need to earn medals. Medals are awarded for great performance in competitions or by the number of upvotes on notebooks (similar to GitHub stars).

We added the tiers information (Novice, Contributor, Expert, Master, Grand Master) to all the studied notebooks. 
We only considered the Code and Competition tracks. The Dataset track displays users' performance on the Dataset part of Kaggle, where users can build, share, and use datasets. 
The dataset track is not relevant to this study, considering that we are not looking to measure the quality of datasets but the quality of the ML pipelines. 

Since users can reach new tiers with time, we considered the current progression stage of the users at the time they authored the notebooks.
By doing that, we can have notebooks authored by the same users with different expertise levels. 

Notebooks submitted in a competition can be authored by a team of users ($85\%$ of our notebooks are authored by a single user. Our largest team is 23 users). 
In this scenario, we chose to retain the expertise level of the user with the highest expertise level among team members. 
Our rationale is based on the assumption that the most experienced author is likely to have reviewed the code and corrected errors or bad practices before submission. 
We also argue that averaging expertise levels (e.g., a Novice and a Grand Master) is inaccurate, as a team’s output is more likely to align with the standards of its most skilled member rather than gravitating toward a median "Expert" level.
The impact of this choice is mitigated by the fact that $85\%$ of our notebooks are single-authored.

\subsubsection{Rule Violation Density} 
Larger notebooks are likely to contain more violations than smaller ones.
To account for notebook size, we counted the number of violations that we normalized to report the violation density (number of violations per 100 lines of code)~\cite{boog08a}. 
\[
density = (\frac{totalViolations}{SLOC}) * 100
\]
During the conversion step (Section~\ref{para:nb_convertion}), the tool used to transform notebooks into Python scripts inserted commented lines into the source code to represent cell metadata and non-Python elements (such as notebook magic commands). 
To prevent these additions from biasing the violation density metrics, we used the Source Lines of Code (SLOC) metric, which accounts only for executable Python lines and excludes comments. 
We verified that none of the analysis rules used in this study target commented code, ensuring that this choice does not impact our findings. 

For both Pylint and SonarQube, the rule violations are categorized by severity.
In Pylint, violations are categorized as \textbf{Refactor}, \textbf{Convention}, \textbf{Warning} or \textbf{Error}\footnote{\url{https://pylint.readthedocs.io/en/stable/user_guide/messages/messages_overview.html}}. 
In SonarQube, violations are categorized as \textbf{Info}, \textbf{Minor}, \textbf{Major}, \textbf{Critical} or \textbf{Blocker}\footnote{\url{https://docs.sonarsource.com/sonarqube-community-build/instance-administration/analysis-functions/instance-mode/mqr-mode}}
We computed the density per severity and on the total number of violations for each tool.
We added these density metrics to our dataset. 

\subsubsection{Quality score} 
For each notebook, we added two quality scores, one displaying the Python quality and one displaying the ML quality.
For Python, we used the quality score given by Pylint.  
\begin{equation*}
    \text{score} = \max \left( 0, 
    10.0 - \dfrac{5 \times E + W + R + C}{S} \times 10 
    \right)
\end{equation*}

Where 
\begin{itemize}
    \item $E$ (\textit{Errors}): number of Error violations (weight: 5);
    \item $W$ (\textit{Warnings}): number of Warning violations ;
    \item $R$ (\textit{Refactor}): number of Refactoring ;
    \item $C$ (\textit{Convention}): number of Style violations ;
    \item $S$ (\textit{Statements}): number of Statements.
\end{itemize}

For the ML score, we adapted the Pylint formula to SonarQube violations by weighting Blocker issues equivalently to Pylint Errors due to their comparable severity. 
To avoid further arbitrary choices, we left the remaining severities unweighted. 
We discuss the implications of this choice in the Threats to Validity section~[\ref{sec:threats-construct}].
 
\begin{equation*}
    \text{score} = \max \left( 0, 
    10.0 - \dfrac{5 \times B + C + MA + MI + I}{S} \times 10 
    \right)
\end{equation*}

Where 
\begin{itemize}
    \item $B$ (\textit{Blocker}): number of Blocker smells (weight: 5);
    \item $C$ (\textit{Critical}): number of Critical smells;
    \item $MA$ (\textit{Major}): number of Major smells;
    \item $MI$ (\textit{Minor}): number of Minor smells;
    \item $I$ (\textit{Information}) : number of Information;
    \item $S$ (\textit{Statements}): number of Statements (given by \textit{Radon} as LLOC).
\end{itemize}

\subsubsection{Machine Learning performance standardized score.}
\label{z-score}
Due to task heterogeneity in competitions, raw performance scores are not comparable across groups. 
Competition tasks vary a lot (classification, prediction, generation,\dots), and the method to compute scores also varies, even in the same type of tasks (F1, Root Mean Squared Error,\dots).
The score range and polarity also change from competition to competition. 
Sometimes higher is better, and sometimes lower is better, depending on the score metric.   
We therefore apply a within-group $z$-score standardization \cite{andrade2021zscores}, which captures individuals’ relative performance within their group.
\[
Z = \frac{X - \mu}{\sigma}
\]
Where $Z$ is the z-score, $X$ is the observed value, $\mu$ is the mean, and $\sigma$ is the standard deviation.
For each competition, we aligned the score polarity to higher is better by multiplying the score by minus one when the polarity was in the other direction. 
Then we computed the mean $\mu$ and the standard deviation $\sigma$ using all the scores present in metadata (even if the notebooks were not public, we can still access the metadata containing the score). 
For $32,054$ notebooks, we were unable to calculate the z-score (because we had no performance score or competition standard deviation) so we dropped them, leaving $265,363$ notebooks the final number of notebooks we used for this study. 

This standardization does not aim to make performance scores comparable across the different competitions but rather to express each individual’s relative standing within their own competition. 
It is this relative standing that becomes comparable across our entire dataset.
We consider this approach sufficient for comparing notebooks across diverse competitions and ML tasks. 
A notebook that ranks among the top performers in its specific competition will yield a positive Z-score, quantifying the extent to which it outperforms the group mean. 
Conversely, a notebook performing near the mean will result in a Z-score close to zero (or negative when the notebook's score is below the mean), allowing for comparison of relative performances across our entire dataset.

\subsection{Quality Rules Selection} 

Pylint and SonarQube were originally designed for traditional software.
When applied to notebooks they may generate noise and false positives due to notebook-specific features.
We excluded rules for Pylint and customized SonarQube to focus on ML specific practices.

\paragraph{Pylint rules}
\label{sec:quality_rules:pylint}
We excluded specific Pylint rules that are not relevant in our context regarding the Kaggle data and the notebook format:
\begin{itemize}
\item \textbf{File naming conventions:} Kaggle renames notebooks using numerical unique identifiers, which we preserved in our dataset. Therefore, enforcing standard Python file-naming conventions is irrelevant.
\item \textbf{Pointless statements and Expression not assigned:} In the context of computational notebooks, it is standard practice to place a variable or object on the last line of a cell to trigger its visual representation (e.g., displaying a DataFrame or a plot). Pylint typically flags these as 'pointless', but since they are intentional in this medium, we ignored them to avoid noise in our quality metrics. The same thing happens for expression statements at the end of a notebook cell, where the result of the expression is automatically displayed. This exclusion is consistent with prior studies on notebook quality~\cite{van21a,quaranta2022a}, which similarly dismiss these rules, as they are recognized as false positives in the notebook medium.
\item \textbf{Dependency-aware rules:} Some rules (e.g., import-error, no-member) require the execution environment to check imports and object members. Because this environment is unavailable in our pipeline, we deactivated these checks by applying the \texttt{--ignored-modules=*} option (see Section~\ref{para:nb_convertion}).
\end{itemize}

\paragraph{SonarQube rules}
For SonarQube, we configured a quality profile containing only rules related to data science, machine learning, and ML libraries.
The profile comprises 34 rules detailed in Appendix~\ref{detail_sonar_selected_rules}.

\section{Study Design}
\label{sec:study_design}

This section presents our study design. 
We first describe the statistical framework common to all research questions (\ref{statistical_framework}). 
We then identify confounding variables, empirically assess their impact, and derive the two analysis samples used throughout the study (\ref{coufounding_variables}). 
Sections \ref{methodo_rq1} to \ref{methodo_rq3} describe the methodology specific to each research question. 
Section~\ref{consolidated_power_analysis} consolidates the power analyses for all tests.

\subsection{Statistical framework}
\label{statistical_framework}

\paragraph{Statistical tests}
To avoid making restrictive assumptions about the distribution of our data, we exclusively employed non-parametric statistical tests. 
We used Spearman’s rank correlation~\cite{Mukaka2012} to evaluate relationships between variables and the Mann-Whitney-Wilcoxon U test~\cite{Fay2010} to compare independent groups. 
Using those tests ensures the robustness of our results, even in the presence of non-normally distributed data or outliers.

\paragraph{Effect size}
\label{effect_size_threshould}
We followed the revised Cohen's guidelines~\cite{fund19a} and used the threshold displayed in Table~\ref{tab:effect-thresholds}  for Spearman's $\rho$ and for the Rank-Biserial 
Correlation Coefficient ($R_{rb}$).
We restrict our analysis to the Small, Medium, and Large ranges, and disregard
effects falling outside of them. Specifically, we treat any effect below the
Small threshold as Negligible, and we do not consider Very Small~\cite{fund19a}
effects, since the conditions under which such effects become meaningful do not
arise in our setting. Likewise, we do not differentiate Large and Very Large effects.

\begin{table}[H]
\centering
\begin{tabular}{lcc}
\hline
\textbf{Effect size} & \textbf{Spearman $\rho$} & \textbf{$R_{rb}$} \\
\hline
Small      & $0.10 \le |\rho| < 0.20$ & $0.10 \le |R_{rb}| < 0.20$ \\
Medium     & $0.20 \le |\rho| < 0.30$ & $0.20 \le |R_{rb}| < 0.30$ \\
Large      & $0.30 \le |\rho| < 0.40$ & $0.30 \le |R_{rb}| < 0.40$ \\
\hline
\end{tabular}
\caption{Effect size interpretation thresholds used in this study.} 
\label{tab:effect-thresholds}
\end{table}

For the Mann-Whitney-Wilcoxon test, we additionally report the Common Language
Effect Size (CLES)~\cite{vargha2000critique}. Since the test itself only test the null hypothesis that random values from two groups have the same distribution, the CLES complements it by
quantifying the magnitude of that difference as the probability that a randomly selected observation from one group exceeds a randomly selected observation from the other.

\paragraph{Power}
For each statistical test, we conducted an \textit{a priori} sensitivity test to compute the minimum detectable effect size.
Using G$^*$Power~\cite{faul2007g}\footnote{Version 3.1.9.6}, 
we followed recent methodological recommendations~\cite{lakens2022sample} by aiming for a statistical power of $0.95$ ($0.99$ for the assessment of confounding variables) and a significance level ($\alpha$) of $0.005$~\cite{benjamin2018redefine}. 
As highlighted by Lakens et al.~\cite{lakens2022sample}, the conventional $80\%$ power threshold is largely arbitrary, and researchers should aim for higher statistical power whenever resource constraints allow. 
Given the substantial scale of our dataset, we were able to adopt these more conservative thresholds, thereby augmenting the robustness of our statistical inferences.

\paragraph{Effect size convertion formula}
\label{effect_convertion_formula}
In some of our tests the effects are reported in a $d$ type effect size (based on the magnitude of mean difference), and we need to convert them to an $r$ type effect size~\cite{bore09a,aaro98a} (on the magnitude of correlation) to use our effect size thresholds defined in~\ref{effect_size_threshould}.
We used the Cohen's conversion formula adapted for groups of different size~\cite{bore09a,aaro98a}.

$$r = \frac{d}{\sqrt{d^2 + a}}$$

$$a = \frac{(n_1 + n_2)^2}{n_1 n_2} $$

Where 
\begin{itemize}
    \item $d$ : Effect size $d$;
    \item $r$ : Effect size $r$;
    \item $a$ : Correction factor for the case where $n_1 \neq n_2$;
    \item $n$ : Sample size.
\end{itemize}

\subsection{Confounding Variables}
\label{coufounding_variables}
Section~\ref{dataset_building} allowed us to build a dataset containing 265,363 notebooks with metadata, scores, and detailed violations.
To ensure the validity of our statistical analyses, we must account for several confounding variables inherent to the competition's structure.

\paragraph{Selection bias.}
Teams can usually select up to two notebooks for final grading. 
We need to determine if these selected notebooks differ in performance from the intermediate drafts (non-selected notebooks) to mitigate potential bias from the non-selected notebooks.

\paragraph{Post-deadline submission bias.}
Notebooks submitted after the competition deadline may benefit from knowledge spillover.
Those notebooks submitted sometimes years after the competition may benefit from the public release of winning strategies (disclosed after the end of the competition) or from recent advances in the ML domain of the competition.
Such a benefit might bias the correlation with the performance or practices of the given notebooks.
Those notebooks may also be biased due to the absence of time constraints. 
Kaggle competitions usually last a few weeks or months; during this time limit, practitioners may prioritize ML performance over code quality. 
Notebooks submitted after the end of the competition are not subject to these time constraints.

\paragraph{Intra-team dependency bias.}
Teams (or individual authors) frequently iterate by cloning an initial notebook, modifying it, and submitting it, resulting in multiple similar notebooks with minor adjustments.  
This iterative nature of submissions introduces an intra-team dependency. 
Treating every submission as an independent observation would lead to pseudo-replication~\cite{hurl84a}, where the specific coding habits or errors of highly active teams are over-represented, thus biasing the results. 
Unlike the two previous biases, this one need not be measured: it can be neutralized by construction, by keeping a single notebook per team and per competition, so that every observation is independent (Section~\ref{final_sampling}).

\paragraph{Confounding by skill.} 
\label{coufounding_variables:skill}
More capable practitioners may simultaneously write better ML code and obtain better results, with no direct link between the two.
To account for this possible confounding, we will stratify the samples used in our tests by expertise and competition type. 

\hfill \break
Whereas the intra-team dependency is removed directly by our sampling design, the post-deadline submission bias and the selection bias cannot be eliminated by construction.
We therefore evaluated their impact (Section~\ref{coufounding_methodo}) to determine whether they threaten the validity of our analyses, and we used these results to define the final datasets used in this study (Section~\ref{final_sampling}).
We took into account the skill confounder in the design of RQ2 and RQ3. 

\subsubsection{Statistical Methodology}
\label{coufounding_methodo}
To evaluate the impact of the selection bias and the post-deadline submission bias, we performed two statistical tests.
First, we tried to differentiate the notebooks submitted during competition time from those submitted after the deadline based on the performance score. 
Then, we tested to differentiate between both selected by users and not selected by users groups over the performance score.

\paragraph{Power analysis.}
\label{sampling_groups_power_analysis}
For both tests, we used a Mann-Whitney U test. 
We aimed for a statistical power of $0.99$ and an alpha error ($\alpha$) of $0.005$. 
For the first test, considering both user-selected notebooks group ($n=11,918$) and non-selected notebooks group ($n=253,445$), parameters and results are shown in Table~\ref{tab:gpower_sensitivity_wilcoxon_selected}. 

For the second test, considering the notebooks submitted before the competition deadline ($n=214,294$) and those submitted after ($n=51,069$), parameters and results are shown in Table~\ref{tab:gpower_sensitivity_wilcoxon_deadline}. 

As outlined in section~\ref{statistical_framework}, the sensitivity analysis yields the minimal detectable effect in Cohen's $d$, which we converted to an $r$ effect for consistency.
We used the conversion formula~\ref{effect_convertion_formula} to convert the results to an $r$ type effect size.
For the test between user-selected and non-selected notebooks, we found a sensitivity $d = 0.049$.
For the test between notebooks submitted before and after the competition deadline, we found a sensitivity $d = 0.026$.
Following the conversion formula, both comparisons reached a minimal detectable effect of $r=0.01$, a negligible effect under the thresholds of Table~\ref{tab:effect-thresholds}.
The complete calculus can be found in Appendix~\ref{effect_size_convertion_selected_not_selected}.

\paragraph{Results.}
The results are summarized in Table~\ref{tab:mann_whitney_coufunding_selection} and \ref{tab:mann_whitney_coufunding_deadline}. 
For the comparison between selected and non-selected notebooks, we observed an $R_{rb} = 0.135$ regarding performance scores.

Under the thresholds of Table~\ref{tab:effect-thresholds}, an $R_{rb} = 0.135$ indicates a small effect, further supported by a Common Language Effect Size ($CLES$)~\cite{vargha2000critique} of $0.57$, meaning a selected notebook has only a $57\%$ chance of outperforming a non-selected one. 
Regarding the number of Pylint violations and the number of ML violations, we observed negligible effects.
These results suggest that while selected notebooks exhibit slightly higher performance, they do not differ from non-selected ones regarding the number of Pylint or ML violations.

Regarding the post-deadline submission bias, the differences between notebooks submitted before and after the deadline were negligible for all observed variables (performance ($R_{rb} = -0.037$), Pylint violations ($R_{rb} = -0.01$) and ML violations ($R_{rb} = -0.05$)). 
These results suggest that submission timing does not significantly impact the code quality or the performance of the notebooks in our dataset.

\begin{table}[h]
\centering
\caption{Wilcoxon-Mann-Whitney test between user selected notebooks ($n_1=11,918$) and non-selected notebooks ($n_2=253,445$). $CI=95$}
\label{tab:mann_whitney_coufunding_selection}
\begin{tabular}{lrrrr}
\textbf{Variable} & \textbf{($R_{rb}$)} & $CI=95$ & \textbf{$p$-value} & \textbf{Effect} \\ 
Performance score           & \textbf{0.135} & $[0.125,0.146]$ & $6.32e^{-138}$ & Small \\
Pylint violations (norm.)   & -0.05 & $[-0.06,-0.04]$ & $1.9e^{-20}$ & Negligible  \\
ML violations (norm.)   & 0.052 & $[0.041,0.062]$ & $1.35e^{-21}$ & Negligible  \\ 
\end{tabular}
\end{table}

\begin{table}[h]
\centering
\caption{Wilcoxon-Mann-Whitney test between notebooks submitted before the end of the competition ($n_1=214,294$) and after ($n_2=51,069$). $CI=95$}
\label{tab:mann_whitney_coufunding_deadline}
\begin{tabular}{lrrrr}
\textbf{Variable} & \textbf{($R_{rb}$)} & $CI=95$ & \textbf{$p$-value} & \textbf{Effect} \\ 
Performance score           & -0.037 & $[-0.04,-0.03]$ & $4.52e^{-38}$ & Negligible \\
Pylint violations (norm.)   & -0.01 & $[-0.019,-0.008]$ & $4.14e^{-07}$ & Negligible  \\
ML violations (norm.)   & -0.05 & $[-0.05,-0.04]$ & $8.47e^{-63}$ & Negligible  \\ 
\end{tabular}
\end{table}

\subsubsection{Final Sampling}
\label{final_sampling}
Based on these findings, we defined two distinct groups for our empirical analysis.

\paragraph{The Full Dataset ($N = 265,363$).}
\label{full_dataset}
We used the complete set of notebooks to answer RQ1. 
For RQ1, popularity (upvotes) is a notebook-specific metric. 
Iterative versions of the same notebook within the same team do not share the same popularity, making the full granularity necessary. 

\paragraph{The Best-Per-Team Dataset ($N = 45,443$).}
\label{best_per_team_dataset}
To answer RQ2 and RQ3, we retained only the notebook with the highest performance score for each team per competition. 
This filtering is crucial to mitigate intra-team dependency and avoid pseudo-replication, ensuring that each data point is independent.

Our statistical tests revealed that the selection bias is small ($R_{rb} = 0.135$). 
Notebooks in the selected group achieve a higher performance score in approximately 57 percent of pairwise comparisons with non-selected notebooks. 
The post-deadline submission bias is negligible. 
Consequently, this sampling strategy (keeping only the notebook with the highest performance score for each team per competition) allows us to capture the notebooks driving the observed performance differences, while still including submissions made after the competitions ended.
By controlling for these structural biases, our sampling strategy ensures that the subsequent analyses will not be biased by dataset artifacts.

\subsection{RQ1. Is there a correlation between notebook popularity and code quality?}
\label{methodo_rq1}

\paragraph{Dataset.} We used the \textit{Full Dataset} ($n=265,363$) (see section~\ref{full_dataset}).

\paragraph{Popularity metric} We used the number of upvotes as a proxy for notebook popularity. 
The upvote mechanism on Kaggle is similar to the one of Stack Overflow.
Users can upvote individual notebooks, and each notebook, including copies of original notebooks maintains its own unique upvote count.
We considered alternative metrics, such as the number of views or the number of copies, but found them less reliable for this study. 
Specifically, view counts on Kaggle are inherited and aggregated on the original notebook. 
Assigning the same value to all subsequent copies would have introduced significant bias in our analysis. 
Similarly, the total number of copies can be misleading.
Authors often iterate on their own work by creating multiple clones, thereby artificially inflating the count. 
While we could have filtered for copies created exclusively by external users (not one of the original authors), we concluded that upvotes provide a more direct and less noisy signal of the popularity. 

\paragraph{Variables.} We observed the correlation between the number of upvotes with the Pylint quality score and the ML quality score

\paragraph{Statistical test.} We assess each correlation with a two-sided Spearman test.

\paragraph{Power} The minimal detectable correlation $\rho = 0.00819$ (see Table~\ref{tab:gpower_sensitivity}).
This corresponds to negligible effect under the interpretation thresholds of Table~\ref{tab:effect-thresholds}.

\subsection{RQ2. How do violations of best practices relate to machine learning performance?}

\paragraph{Dataset.} We used the \textit{Best-Per-Team Dataset} ($n = 45,443$), which retains only the highest-performing notebook per team per competition (see section~\ref{best_per_team_dataset}). 

\paragraph{Variables.} For each notebook, we correlated the ML 
performance (z-score) with four code quality variables:
\begin{itemize}
    \item the Pylint quality score and the ML quality score,
    \item the normalized number of Pylint violations and the 
    normalized number of ML violations (violations density).
\end{itemize}
We report both score-based and density-based variables to ensure 
that our conclusions do not depend on the weighting choices made in 
the score formulas.

\paragraph{Statistical test.} We assessed each correlation with a two-sided Spearman test.

\paragraph{Analysis levels.} We analyzed RQ2 at four nested levels 
of granularity. We first measured the correlation 
\emph{(a)} on the entire Best-Per-Team Dataset, to capture the 
overall trend. 
We then disaggregated the data to verify that the overall trend is not driven by contextual factors:
\emph{(b)} stratified by competition type, 
\emph{(c)} stratified by author expertise, and 
\emph{(d)} by the intersection of competition type and expertise.

\paragraph{Competition types.} 
\label{sec:rq2_methodo:competition_type}
We focused on the three competition 
types that satisfy our statistical power requirements: 
\emph{Playground} ($n = 20\,965$), \emph{Featured} ($n = 14\,802$), 
and \emph{Research} ($n = 7\,813$). We excluded \emph{Community} 
($n = 1\,601$), \emph{Recruitment} ($n = 228$), and 
\emph{Analytics} ($n = 34$) competitions, whose sample sizes are 
too small to detect even small effects with our target power.

\paragraph{Expertise groups}
\label{rq2_design_expertise_groups}
As presented in~\ref{users_expertise}, Kaggle users have an expertise tier on \textit{Code}, \textit{Competition} or both.
Tiers span from \textbf{Novice} to \textbf{Grand Master} (\textbf{Novice}, \textbf{Contributor}, \textbf{Expert}, \textbf{Master}, and \textbf{Grand Master}). 
For our study we considered two groups, the experts and the non-experts.
Experts are users with one the three following level: Expert, Master, and Grand Master.
Non-expert are Novice or Contributor users.
We chose to merge the expertise levels that require performance achievement or community validation into the Experts groups.
While this binary classification sacrifices the granularity of the five-tier system, it ensures sufficient group sizes for statistical testing and reflects the distinction between users who have received community or competitive validation and those who have not.

\paragraph{Power.} Sample sizes and minimal detectable correlations 
for all configurations of this RQ are consolidated in Table~\ref{tab:power-consolidated}. 
Across all configurations, the minimal detectable $\rho$ ranges from $0.021$ (overall) to $0.135$ (Research competitions, competition experts), all corresponding to negligible to small effects under the interpretation thresholds of Table~\ref{tab:effect-thresholds}.

\subsection{RQ3. Do expert and non-expert users differ in their violation of best practices?}
\label{methodo_rq3}

\paragraph{Dataset.} We used the \textit{Best-Per-Team Dataset} ($n = 45\,443$) (see section~\ref{best_per_team_dataset}). 


\paragraph{Variables.} Whereas RQ2 correlated quality with performance, RQ3 compares the distribution of code quality between expert and non-expert groups. 
For each group, we compare the normalized number of violations, broken down by tool and by severity:
\begin{itemize}
    \item the total normalized Pylint violations and their breakdown 
    by severity (Error, Warning, Convention, Refactor);
    \item the total normalized ML violations and their breakdown by 
    severity (Blocker, Critical, Major, Minor).
\end{itemize}
We also include the performance (z-score) in the comparison. 
Analyzing violations at the severity level, rather than only in aggregate, allows us to detect whether differences between groups are driven by specific categories of violations.

\paragraph{Statistical test.} We compared the two groups with a two-sided Wilcoxon-Mann-Whitney U test. 

\paragraph{Expertise groups}
We used the same expert and non-expert groups as in the previous question~\ref{rq2_design_expertise_groups}.

\paragraph{Power.} This RQ relied on the same group sizes as the author expertise level of RQ2.
The competition tracks compare $3\,620$ experts against $41\,823$ non-experts, and the code tracks compare $6\,529$ experts against $38\,914$ non-experts. 
As outlined in section~\ref{statistical_framework}, the sensitivity analysis yields the minimal detectable effect in Cohen's $d$, which we convert to $r$ for consistency. 
Both comparisons reach a minimal detectable effect of $r = 0.021$ (see Table~\ref{tab:power-consolidated}), a negligible effect under the thresholds of Table~\ref{tab:effect-thresholds}.

\subsection{Consolidated Power Analysis}
\label{consolidated_power_analysis}

All tests use $\alpha = 0.005$. Power is set to $0.99$ for confounder assessments and $0.95$ for research questions. 
MWW = Wilcoxon-Mann-Whitney U test. 
For MWW tests, the minimal detectable effect is reported both as Cohen's $d$ (G*Power output) and as $r$ after conversion (see Section~\ref{effect_convertion_formula}). 
Detailed G*Power parameters are reported in Appendix~\ref{g_power_appendix}.

\begin{table}[!ht]
\caption{Consolidated sensitivity power analysis for all statistical tests 
performed in this study.}
\label{tab:power-consolidated}

\centering
\setlength{\tabcolsep}{4pt}
\renewcommand{\arraystretch}{1.15}
\resizebox{\textwidth}{!}{%
\begin{tabular}{llrrcc}
\hline
\textbf{Analysis} & \textbf{Test} & \textbf{Sample size} & \textbf{Power} 
& \textbf{Min. detectable effect} & \textbf{Interpretation} \\
\hline
\multicolumn{6}{l}{\textit{Confounder assessment}} \\
\quad Selection bias            & MWW      & (11\,918,\ 253\,445) & 0.99 & $d = 0.049 \rightarrow r = 0.01$ & Negligible \\
\quad Post-deadline submission  & MWW      & (214\,294,\ 51\,069) & 0.99 & $d = 0.026 \rightarrow r = 0.01$ & Negligible \\
\hline
\multicolumn{6}{l}{\textit{RQ1 — Popularity vs code quality}} \\
\quad Overall                   & Spearman & 265\,363 & 0.95 & $\rho = 0.0082$ & Negligible \\
\hline
\multicolumn{6}{l}{\textit{RQ2 — Best practices vs ML performance}} \\
\quad Overall                   & Spearman & 45\,443 & 0.95 & $\rho = 0.021$ & Negligible \\
\quad By competition: Featured  & Spearman & 14\,802 & 0.95 & $\rho = 0.037$ & Negligible \\
\quad By competition: Playground & Spearman & 20\,965 & 0.95 & $\rho = 0.031$ & Negligible \\
\quad By competition: Research  & Spearman & 7\,813  & 0.95 & $\rho = 0.050$ & Negligible \\
\quad By expertise: Competition exp.       & Spearman & 3\,620  & 0.95 & $\rho = 0.074$ & Negligible \\
\quad By expertise: Competition non-exp.   & Spearman & 41\,823 & 0.95 & $\rho = 0.022$ & Negligible \\
\quad By expertise: Code exp.              & Spearman & 6\,529  & 0.95 & $\rho = 0.055$ & Negligible \\
\quad By expertise: Code non-exp.          & Spearman & 38\,914 & 0.95 & $\rho = 0.023$ & Negligible \\
\multicolumn{6}{l}{\textit{By competition $\times$ Competition expertise}} \\
\quad Comp. exp.: Featured       & Spearman & 1\,862  & 0.95 & $\rho = 0.098$ & Negligible \\
\quad Comp. exp.: Playground     & Spearman & 541     & 0.95 & $\rho = 0.180$ & Small \\
\quad Comp. exp.: Research       & Spearman & 1\,117  & 0.95 & $\rho = 0.126$ & Small \\
\quad Comp. non-exp.: Featured   & Spearman & 12\,940 & 0.95 & $\rho = 0.037$ & Negligible \\
\quad Comp. non-exp.: Playground & Spearman & 20\,424 & 0.95 & $\rho = 0.029$ & Negligible \\
\quad Comp. non-exp.: Research   & Spearman & 6\,696  & 0.95 & $\rho = 0.051$ & Negligible \\
\multicolumn{6}{l}{\textit{By competition $\times$ Code expertise}} \\
\quad Code exp.: Featured        & Spearman & 1\,908  & 0.95 & $\rho = 0.096$ & Negligible \\
\quad Code exp.: Playground      & Spearman & 3\,509  & 0.95 & $\rho = 0.071$ & Negligible \\
\quad Code exp.: Research        & Spearman & 970     & 0.95 & $\rho = 0.135$ & Small \\
\quad Code non-exp.: Featured    & Spearman & 12\,894 & 0.95 & $\rho = 0.037$ & Negligible \\
\quad Code non-exp.: Playground  & Spearman & 17\,456 & 0.95 & $\rho = 0.032$ & Negligible \\
\quad Code non-exp.: Research    & Spearman & 6\,843  & 0.95 & $\rho = 0.050$ & Negligible \\
\hline
\multicolumn{6}{l}{\textit{RQ3 — Expert vs non-expert violations}} \\
\quad Competition expertise     & MWW      & (3\,620,\ 41\,823)  & 0.95 & $d = 0.079 \rightarrow r = 0.021$ & Negligible \\
\quad Code expertise            & MWW      & (6\,529,\ 38\,914)  & 0.95 & $d = 0.061 \rightarrow r = 0.021$ & Negligible \\
\hline
\end{tabular}
}
\end{table}

\section{Results}
\label{sec:results}
In this section, we present our results. All reported $p$-values are exact.

\subsection{RQ1. Is there a correlation between notebook popularity and code quality?}

\begin{tcolorbox}[colback=gray!10, colframe=gray!50, arc=2pt, boxrule=0.8pt]
\paragraph{RQ1 Summary.} There is a negligible correlation between the popularity (number of upvotes a notebook receives) and its quality score. 
\textbf{The popularity of a notebook does not imply that it contains quality code.
Therefore, users should not rely on popularity when searching for notebooks exhibiting quality code}.
\end{tcolorbox}
\noindent
Using the \textit{Full Dataset} ($n=265,363$), we looked at the correlation between Python and ML quality score with the popularity of the notebook (number of upvotes).

\paragraph{Results}
Results for Spearman correlation tests between the number of upvotes and the Pylint quality and between the number of upvotes and the ML quality score are shown in Table~\ref{tab:results_summary}.
Both correlations are above the minimal detectable correlation, which is consistent with the required statistical power. 
P-values are significant for both tests ($p < 0.005$). 
Since we are not comparing multiple correlations, we can directly use the Spearman $\rho$ as a measure of the effect size.
Following the interpretation thresholds of Table~\ref{tab:effect-thresholds},
$|\rho| < 0.1$ is negligible, therefore both observed correlations are negligible.



\begin{table}[h!]
\centering
\caption{Spearman correlation between quality scores and up-votes ($n=265,363$, minimal detectable correlation $\rho=0.0082$).}
\label{tab:results_summary}
\begin{tabular}{lccc}
\textbf{Variable correlated to up-votes}  & \textbf{Observed $\rho$} & \textbf{$p$-value} & \textbf{Effect} \\ 
Pylint quality score  &  $0.015$ &  $2.64e^{-15}$ & Negligible \\ 
ML quality score      & $-0.079$ & $0.0$ & Negligible \\ 
Performance      & $0.0007$ & $0.7315$ & NS \\ 
\end{tabular}
\end{table}

\subsection{RQ2. How do violations of best practices relate to machine learning performance?}

\begin{tcolorbox}[colback=gray!10, colframe=gray!50, arc=2pt, boxrule=0.8pt]
\paragraph{RQ2 Summary.} 
We observed a significant negative correlation between ML violations and performance: fewer violations relate to higher performance. 
This effect is small but consistent across expertise levels and competition types. 
In contrast, general Python quality (Pylint) has a negligible impact on performance. 
\textbf{These findings indicate that general Python quality is unrelated to performance, neither hindering nor helping it. However, adopting ML best practices is associated with superior performance outcomes, even if it does not guarantee them.}
\end{tcolorbox}
\noindent
For this research question, we used the \textit{Best-Per-Team Dataset} ($n = 45,443$) consisting of the highest-performing notebook per team per competition. 

\subsubsection{Correlation between performance score and quality scores}

\paragraph{Results.}
Table \ref{tab:corr_perf_zscore_g2} summarizes the Spearman correlation between ML performance (z-score) and quality scores. 
We observe a significant but negligible negative correlation between the Pylint quality score and performance ($\rho = -0.0795$).
We observe a significant negligible positive correlation between the number of Pylint violations and the performance ($\rho = 0.053$).
The correlation between ML quality score and performance shows a significant small effect ($\rho = 0.120$). 
This indicates a positive relationship: notebooks with higher ML quality scores (better adherence to ML best practices) tend to achieve higher performance on the leaderboard.
We observe the same relation between the number of ML violations and the performance, where we have a significant small negative effect ($\rho = -0.132$).
Meaning that notebooks with fewer ML violations tend to achieve better performances.

\begin{table}[h!]
\centering
\caption{Spearman between Performance (z-score) and quality scores ($n=45,443$). ($\rho_{crit}=0,021$)}
\label{tab:corr_perf_zscore_g2}
\begin{tabular}{lccc}
\textbf{Variable correlated to performance} & \textbf{Observed $\rho$} & \textbf{$p$-value} & \textbf{Effect} \\ 
Pylint Quality Score  & $-0.0795$ & $1.18 \times 10^{-64}$ & Negligible \\ 
ML Quality Score      & $0.120$  & $6.19 \times 10^{-146}$ & Small \\ 
Pylint nb Violations (norm.)  & $0.053$ & $1.51 \times 10^{-29}$ & Negligible \\ 
ML nb Violations (norm.)     & $-0.132$ & $3.25 \times 10^{-176}$ & Small \\ 
\end{tabular}
\end{table}

\subsubsection{Analysis by Contextual Factors}
The global trends observed in the \textit{Best-Per-Team Dataset} might be influenced by the specific context of the competition or the expertise of the notebook's author. 
To ensure the robustness of our findings, we disaggregated the data by competition type and user expertise.

\paragraph{Analysis with stratification by competition type.}
In this section, we looked at the correlation between notebook performance and the number of violations.
We observed the correlation on each type of competition (Playground, Featured, Research, \dots) independently. 

\paragraph{Results.}
The results (Table~\ref{tab:competition_types_final}) show that the number of Pylint violations has a non-significant or negligible correlation with performance across all competition types. 
In contrast, the number of ML-specific violations consistently shows a significant negative correlation with performance. This effect is most pronounced in Research competitions ($\rho = -0.193$, small effect).
The effect remains significant but smaller in Playground and Featured competitions ($\rho = -0.114$ and $\rho = -0.141$).

\begin{table}[h!]
\centering
\caption{Spearman correlation by competition types (Performance vs nb Violations (norm.))}
\label{tab:competition_types_final}
\begin{tabular}{lcclccc}
\textbf{Competition} & \textbf{$n$} & \textbf{$q_{crit}$} & \textbf{Tool} & \textbf{Observed $q$} & \textbf{$p$-value} & \textbf{Effect} \\ 
\multirow{2}{*}{Featured} & \multirow{2}{*}{14 802} & \multirow{2}{*}{0.037} & Pylint & 0.038 & $3.31e^{-06}$ & Negligible \\ 
 & & & ML & -0.141 & $1.96e^{-67}$ & Small \\  \hline
\multirow{2}{*}{Playground} & \multirow{2}{*}{20 965} & \multirow{2}{*}{0.031} & Pylint & 0.0735 & $1.47e^{-26}$ & Negligible \\ 
 & & & ML & -0.114 & $5.99e^{-62}$ & Small \\  \hline
\multirow{2}{*}{Research} & \multirow{2}{*}{7 813} & \multirow{2}{*}{0.050} & Pylint & 0.0146 & 0.195 & NS \\ 
 & & & ML & -0.193 & $1.54e^{-66}$ & Small \\ 
\end{tabular}
\end{table}

\paragraph{Analysis with stratification by Authors Expertise.}
In this section, as before, we looked at the correlation between notebook performance and the number of violations.
We looked at those correlations by expertise groups over the whole \textit{Best-Per-Team Dataset} and then by splitting the dataset per competition type. 
We distinguished the expertise in competition and the expertise in "code" (see~\ref{users_expertise} and~\ref{rq2_design_expertise_groups}) as they might not depict the same correlation.

\paragraph{Results.}
Results are shown in Table~\ref{tab:results_corr_perf_competition_expert}, \ref{tab:results_corr_perf_competition_non_expert}, \ref{tab:results_corr_perf_code_expert}, \ref{tab:results_corr_perf_code_non_expert}, \ref{tab:competition_types_experts_comp}, \ref{tab:competition_types_non_expert_comp}, \ref{tab:competition_types_expert_code}, and \ref{tab:competition_types_non_expert_code}.
In all configurations, we observed the same trend as previously. 
The correlation between the number of Pylint violations (or Pylint score) is either negligible, non significant, or below our minimal observable correlation for the expert and non-expert, globally, and for all the competition types.

On the complete \textit{Best-Per-Team Dataset}, for both code experts and competition experts, we observe a significant, small negative correlation between ML violations and performance ($\rho = -0.148$ and $\rho = -0.157$ respectively).
For the non-experts in both code and competitions, we observe weaker but still significant negative correlation between the number of ML violations and the performance ($\rho = -0.129$ and $\rho = -0.117$, respectively).

When observed by competition type, this correlation also appears.
The strongest relationship is found among competition experts in Research competitions ($\rho = -0.19$, Table \ref{tab:competition_types_experts_comp}). 
Interestingly, for competition experts in Playground competitions, the correlation for ML violations becomes non-significant ($p=0.404$). 
This is likely due to the smaller sample size ($N=541$) and the more casual nature of these competitions, where experimental "quick-and-dirty" code is more common.

\begin{table}[h!]
\centering
\caption{Spearman between Performance (z-score) and quality scores ($n=3620$) on competition experts. $\rho_{crit}=0.074$}
\label{tab:results_corr_perf_competition_expert}
\begin{tabular}{lccc}
\textbf{Variable correlated to performance} & \textbf{Observed $\rho$} & \textbf{$p$-value} & \textbf{Effect} \\ 
Pylint Quality Score & $-0.0689$ & $3.28e^{-5}$ & - \\ 
ML Quality Score      & $0.149$  & $1.56e^{-19}$ & Small \\ 
Pylint nb Violations (norm.)  & $0.0254$ & $0.126$ & NS \\ 
ML nb Violations (norm.)      & $-0.157$ & $1.67e^{-21}$ & Small \\ 
\end{tabular}
\end{table}

\begin{table}[h!]
\centering
\caption{Spearman between Performance (z-score) and quality scores ($n=41,823$) on competition non-experts. $\rho_{crit}=0.022$}
\label{tab:results_corr_perf_competition_non_expert}
\begin{tabular}{lccc}
\textbf{Variable correlated to performance} & \textbf{Observed $\rho$} & \textbf{$p$-value} & \textbf{Effect} \\ 
Pylint Quality Score & $-0.0769$ & $5.99e^{-56}$ & Negligible \\ 
ML Quality Score      & $0.105$  & $7.28e^{-103}$ & Small \\ 
Pylint nb Violations (norm.) & $0.053$ & $1.09e^{-27}$ & Negligible \\ 
ML nb Violations (norm.)     & $-0.117$ & $4.64e^{-128}$ & Small \\ 
\end{tabular}
\end{table}

\begin{table}[h!]
\centering
\caption{Spearman between Performance (z-score) and quality scores ($n=6529$) on code experts. $\rho_{crit}=0.055$}
\label{tab:results_corr_perf_code_expert}
\begin{tabular}{lccc}
\textbf{Variable correlated to performance} & \textbf{Observed $\rho$} & \textbf{$p$-value} & \textbf{Effect} \\ 
Pylint Quality Score & $-0.072$ & $4.60e^{-9}$ & Negligible \\ 
ML Quality Score     & $0.139$  & $2.02e^{-29}$ & Small \\ 
Pylint nb Violations (norm.) & $0.051$ & $3.99e^{-5}$ & - \\ 
ML nb Violations (norm.)     & $-0.148$ & $3.58e^{-33}$ & Small \\ 
\end{tabular}
\end{table}

\begin{table}[h!]
\centering
\caption{Spearman between Performance (z-score) and quality scores ($n=38,914$) on code non-experts. $\rho_{crit}=0.023$}
\label{tab:results_corr_perf_code_non_expert}
\begin{tabular}{lcccc}
\textbf{Variable correlated to performance} & \textbf{Observed $\rho$} & \textbf{$p$-value} & \textbf{Effect} \\ 
Pylint Quality Score & $-0.08$ & $5.24e^{-57}$ & Negligible \\ 
ML Quality Score     & $0.117$  & $1.96e^{-118}$ & Small \\ 
Pylint nb Violations (norm.) & $0.053$ & $3.27e^{-26}$ & Negligible \\ 
ML nb Violations (norm.)     & $-0.129$ & $7.83e^{-145}$ & Small \\ 
\end{tabular}
\end{table}


\begin{table}[h!]
\centering
\caption{Spearman correlation by competition types on group 2 sample competition expert (Performance vs nb violations norm)}
\label{tab:competition_types_experts_comp}
\begin{tabular}{lcclccc}
\textbf{Competition} & \textbf{N} & \textbf{$\rho_{crit}$} & \textbf{Tool} & \textbf{Observed $\rho$} & \textbf{$p$-value} & \textbf{Effect} \\ 
\multirow{2}{*}{Featured} & \multirow{2}{*}{1862} & \multirow{2}{*}{0.098} & Pylint & 0.017 & 0.456 & NS \\ 
 & & & ML & -0.152 & $4.68e^{-11}$ & Small \\ \hline
\multirow{2}{*}{Playground} & \multirow{2}{*}{541} & \multirow{2}{*}{0.18} & Pylint & 0.1 & $0.019$ & NS \\ 
 & & & ML & 0.0359 & $0.404$ & NS \\  \hline
\multirow{2}{*}{Research} & \multirow{2}{*}{1117} & \multirow{2}{*}{0.126} & Pylint & -0.04 & 0.177 & NS \\ 
 & & & ML & -0.19 & $1.126e^{-10}$ & Small \\ 
\end{tabular}
\end{table}

\begin{table}[h!]
\centering
\caption{Spearman correlation by competition types on group 2 sample competition non-expert (Performance vs nb violations norm)}
\label{tab:competition_types_non_expert_comp}
\begin{tabular}{lcclccc}
\textbf{Competition} & \textbf{N} & \textbf{$\rho_{crit}$} & \textbf{Tool} & \textbf{Observed $\rho$} & \textbf{$p$-value} & \textbf{Effect} \\ 
\multirow{2}{*}{Featured} & \multirow{2}{*}{12940} & \multirow{2}{*}{0.037} & Pylint & 0.0426 & $1.22e^{-6}$ & Negligible \\ 
 & & & ML & -0.126 & $4.13e^{-47}$ & Small \\ \hline
\multirow{2}{*}{Playground} & \multirow{2}{*}{20424} & \multirow{2}{*}{0.029} & Pylint & 0.073 & $1.04e^{-25}$ & Negligible \\ 
 & & & ML & -0.117 & $1.81e^{-63}$ & Small \\ \hline
\multirow{2}{*}{Research} & \multirow{2}{*}{6696} & \multirow{2}{*}{0.051} & Pylint & 0.028 & 0.02 & NS \\ 
 & & & ML & -0.176 & $1.34e^{-47}$ & Small \\ 
\end{tabular}
\end{table}

\begin{table}[h!]
\centering
\caption{Spearman correlation by competition types on group 2 sample code expert (Performance vs nb violations norm)}
\label{tab:competition_types_expert_code}
\begin{tabular}{lcclccc}
\textbf{Competition} & \textbf{N} & \textbf{$\rho_{crit}$} & \textbf{Tool} & \textbf{Observed $\rho$} & \textbf{$p$-value} & \textbf{Effect} \\ 
\multirow{2}{*}{Featured} & \multirow{2}{*}{1908} & \multirow{2}{*}{0.096} & Pylint & 0.015 & 0.50 & NS \\ 
 & & & ML & -0.208 & $4.07e^{-10}$ & Small \\ \hline
\multirow{2}{*}{Playground} & \multirow{2}{*}{3509} & \multirow{2}{*}{0.071} & Pylint & 0.059 & $0.000378$ & - \\ 
 & & & ML & -0.08 & $1.13e^{-6}$ & Negligible \\ \hline
\multirow{2}{*}{Research} & \multirow{2}{*}{970} & \multirow{2}{*}{0.135} & Pylint & 0.034 & 0.284 & NS \\ 
 & & & ML & -0.173 & $6.08e^{-8}$ & Small \\ 
\end{tabular}
\end{table}

\begin{table}[h!]
\centering
\caption{Spearman correlation by competition types on group 2 sample code non-expert (Performance vs nb violations norm)}
\label{tab:competition_types_non_expert_code}
\begin{tabular}{lcclccc}
\textbf{Competition} & \textbf{N} & \textbf{$\rho_{crit}$} & \textbf{Tool} & \textbf{Observed $\rho$} & \textbf{$p$-value} & \textbf{Effect} \\ 
\multirow{2}{*}{Featured} & \multirow{2}{*}{12894} & \multirow{2}{*}{0.037} & Pylint & 0.042 & $1.80e^{-6}$ & Negligible \\ 
 & & & ML & -0.13 & $6.03e^{-51}$ & Small \\ \hline
\multirow{2}{*}{Playground} & \multirow{2}{*}{17456} & \multirow{2}{*}{0.032} & Pylint & 0.076 & $4.71e^{-24}$ & Negligible \\ 
 & & & ML & -0.12 & $1.30e^{-56}$ & Small \\ \hline
\multirow{2}{*}{Research} & \multirow{2}{*}{6843} & \multirow{2}{*}{0.05} & Pylint & 0.013 & 0.289 & NS \\ 
 & & & ML & -0.193 & $1.09e^{-58}$ & Small \\ 
\end{tabular}
\end{table}

\subsection{RQ3. Do expert and non-expert users differ in their violation of best practices?}

\begin{tcolorbox}[colback=gray!10, colframe=gray!50, arc=2pt, boxrule=0.8pt]
\paragraph{RQ3 Summary.} 
When considering competition expertise, experts show better performance than non-experts, while making slightly more Python Error or Refactoring violations and moderately fewer ML violations (section~\ref{comp_expert_results}).
When considering code expertise, experts and non-experts cannot be differentiated (section~\ref{code_expert_results}).
Therefore, \textbf{when searching for quality notebooks, users should not rely on code expertise.
However, users can rely on competition expertise to find notebooks with a better ML quality without guarantees on Python quality}.
\end{tcolorbox}
\noindent
For this question, we used the \textit{Best-Per-Team Dataset} ($n = 45,443$).

\subsubsection{Competition expert and non-expert}
\label{comp_expert_results}
Considering the two groups, the experts and the non-experts in competition. 
We used a two-sided Mann-Whitney U test to compare the distribution of different variables in the groups. 

\paragraph{Results}
Results are shown in Table~\ref{tab:mann_whitney_comp_results}.
In our observations, experts exhibit a significant higher performance scores with a medium effect size ($r_{rb} = 0.224, CLES = 61\%$). 
Regarding code quality, we found a negligible difference in the total number of normalized Pylint violations. However, a granular analysis reveals that experts produce significantly more Pylint errors (small effect, $r_{rb} = 0.159$) and refactor-related issues (medium effect, $r_{rb} = 0.268$).
Experts violate fewer ML-specific rules (significant, medium effect, $r_{rb} = -0.278$). This trend is primarily driven by a lower frequency of "Major" ML violations ($r_{rb} = -0.261$).

\begin{table}[h]
\centering
\caption{Wilcoxon-Mann-Whitney test between competition experts ($n_1=3,620$) non-experts ($n_2=41,823$). $CI=95$}
\label{tab:mann_whitney_comp_results}
\begin{tabular}{lrrrr}
\textbf{Variable} & \textbf{($r_{rb}$)} & $CI=95$ & \textbf{$p$-value} & \textbf{Effect} \\ 
Performance score           & \textbf{0.224} & $[0.205,0.243]$ & $2.95e^{-111}$ & Medium \\
Pylint violations (norm.)   & 0.034 & $[0.014,0.053]$ & $0.0007$ & Negligible  \\
Pylint error (norm.)   & \textbf{0.159} & $[0.140,0.178]$ & $0.394e^{-88}$ & Small  \\
Pylint warning (norm.)   & 0.039 & $[0.02,0.059]$ & $8.284e^{-5}$  & Negligible  \\
Pylint convention (norm.)   & -0.012 & $[-0.032,0.007]$ & $0.214$ & NS \\
Pylint refactor (norm.)   & \textbf{0.268} & $[0.249,0.286]$ & $5.28e^{-182}$ & Medium \\
ML violations (norm.)       & \textbf{-0.278} & $[-0.296,-0.260]$ & $4.09e^{-170}$ & Medium \\ 
ML blocker (norm.)       & $-6.929e^{-05}$ & $[-0.02,0.019]$ & $0.872$ & NS \\ 
ML critical (norm.)       & -0.026 & $[-0.045,-0.006]$ & $8.52e^{-5}$ & Negligible \\ 
ML major (norm.)       & \textbf{-0.261} & $[-0.279,-0.243]$ & $1.93e^{-150}$ & Medium \\ 
ML minor (norm.)       & -0.038 & $[-0.057,-0.018]$ & $1.98e^{-12}$ & Negligible \\ 

\end{tabular}
\end{table}

\subsubsection{Code expert and non-expert}
\label{code_expert_results}
Considering the two groups, expert ($N=6529$) and the non-expert ($N=38,914$) in code. 
We used a two-sided Mann-Whitney U test to compare the distribution of different variables in the groups. 

\paragraph{Results.}
The results are presented in Table~\ref{tab:mann_whitney_code_results}.
We observe that Code experts, unlike the Competition experts, do not exhibit a higher performance score ($r_{rb} = 0.028$). 
Several variables, including Pylint warnings and critical ML violations, did not show statistically significant differences ($p > 0.005$).
For the variables with significative effects. We observed negligible effects.

\begin{table}[h]
\centering
\caption{Wilcoxon-Mann-Whitney test between code experts ($n_1=6529$) non-experts ($n_2=38,914$).}
\label{tab:mann_whitney_code_results}
\begin{tabular}{lrrrrr}
\textbf{Variable} & \textbf{($r_{rb}$)} & $CI=95$ & \textbf{$p$-value} & \textbf{Effect} \\ 
Performance score           & 0.028 & $[0.013,0.043]$ & $0.0003$ & Negligible \\
Pylint violations (norm.)   & -0.034 & $[-0.049,-0.019]$ & $7.96e^{-6}$ & Negligible  \\
Pylint error (norm.)   & 0.052 & $[0.037,0.067]$ & $2.07e^{-17}$ & Negligible  \\
Pylint warning (norm.)   & -0.018 & $[-0.033,-0.003]$ & $0.02$  & NS \\
Pylint convention (norm.)   & -0.045 & $[-0.06,-0.03]$ & $6.98e^{-9}$ & Negligible \\
Pylint refactor (norm.)   & 0.033 & $[0.018,0.048]$ & $3.95e^{-6}$ & Negligible \\
ML violations (norm.)       & -0.059 & $[-0.074,-0.044]$ & $2.89e^{-14}$ & Negligible\\ 
ML blocker (norm.)       & 0.0005 & $[-0.015,0.016]$ & $0.1$ & NS \\ 
ML critical (norm.)       & -0.004 & $[-0.019,0.011]$ & $0.46$ & NS \\ 
ML major (norm.)       & -0.052 & $[-0.067,-0.037]$ & $1.18e^{-11}$ & Negligible\\ 
ML minor (norm.)       & -0.003 & $[-0.019,0.012]$ & $0.39$ & NS \\ 
\end{tabular}
\end{table}

\section{Discussion}
\label{sec:discussion}
 

Our motivation is based on a common but largely untested assumption that ``clean code'' and ``high performance'' are independent factors (Section~\ref{sec:introduction}). 
Our results show that this assumption does not hold uniformly. 
It depends on which notion of code quality is considered.
General Python code quality, as captured by Pylint, is effectively \emph{decoupled} from machine learning performance. 
In contrast, ML-specific practices, as captured by our SonarQube profile, are \emph{consistently coupled} with performance.
The performance-quality dichotomy is therefore neither simply true nor simply false.
It holds for general software-engineering practices, but breaks down for ML-specific ones.

\subsection{General quality and ML-specific quality are different constructs}
\label{sec:discussion:constructs}

Code quality is a complex notion, the two tools we employed capture different facets of a notebook's quality. 
Our measurements confirm an intuition that had not, to our knowledge, been empirically tested before:
general Python code quality has no measurable impact on ML performance, whereas ML-specific quality does. 
This asymmetry is explained by the differences between our two tools.

Pylint flags conventions, errors, and refactoring opportunities: naming, layout,
structural smells, and software engineering concerns that are largely orthogonal to ML (i.e general Python code quality). 
A notebook can be stylistically poor and still implement a sound ML pipeline.
Conversely, adherence to general Python best-practices says nothing about whether the pipeline does not contain data-leakage or misuse of ML API. 

Our SonarQube profile, by design, targets data-science
and ML constructs whose misuse can affect the behaviour of the pipeline itself.

Our observation that these two tools' behaviours diverge is thus not a contradiction but evidence that ``code quality'' is not a single variable in the notebook setting.
General Python violations and ML violations carry different information about the artefact.
We also note that general Python best-practices have no effect on ML performance in either direction.
Respecting them does not come at the expense of performance. 



\subsection{The relationship between ML violations and performance survives our controls}
\label{sec:discussion:causation}


Several of our observations suggest that confounding factors cannot fully explain the effect of ML violations on performance.
First, the negative association between ML violations and performance is still observable when we stratify by author expertise. 
The effect persists for both experts and non-experts in code and in competition (Tables~\ref{tab:results_corr_perf_competition_expert}--\ref{tab:results_corr_perf_competition_non_expert},  and Tables~\ref{tab:results_corr_perf_code_expert}--\ref{tab:results_corr_perf_code_non_expert}). 
If expertise (see Section~\ref{coufounding_variables:skill}) were the only driver, the effect should be highly reduced when observing only non-experts,
which is not what we observe. 
Second, the effect is also stable across competition types, which may differ in domain, difficulty and in the population of participants.
Third, the effects of Pylint violation on performance (which a skill confound should also affect) remain negligible everywhere.
The expertise does not uniformly drag every quality metric in the direction of performance.
This last point is limited since ML practitioners do not necessarily have a software engineering (SE) background and may be ML-experts without extensive knowledge on SE-practices~\cite{simm20a}.
 
While acknowledging that another study design would be required to establish a causal mechanism, we interpret the relationship as a robust association that survives the controls available to us.

\subsection{Two expert populations and the limits of social signals}
\label{sec:discussion:experts} 
RQ3 reveals that ``expert'' is not a single population on Kaggle. Competition
experts perform better than non-experts and violate fewer ML-specific rules. 
On the other hand, they do not produce cleaner Python overall.
The difference in total Pylint violations is negligible, and a
granular view shows they actually have more Error
and Refactor violations. 
Competition experts thus appear to optimise for ML correctness 
while remaining indifferent to general code quality (a pattern consistent with Section~\ref{sec:discussion:constructs}).
Their higher performance score is unsurprising given that competition tiers are themselves earned through strong competition results. 
The more informative finding is the fewer ML violations, which is consistent with RQ2's association between fewer ML violations and higher performance.

Code experts behave differently; they are neither better performers nor cleaner coders, with every comparison displaying a negligible effect. 
This is initially surprising, but follows the RQ1 findings. 
The Kaggle code tier is earned through notebook upvotes, and RQ1 showed that popularity is negligibly correlated with quality and negligibly correlated with performance. 
From our observation, we assume that upvotes and, by extension, Kaggle code tier do not reward ML-practices or SE-practices (general Python code quality) but probably other confounding variables.
This result contrasts with prior work stating that Kaggle users vote for notebooks containing quality code~\cite{drozdova2023a}.   
The internal consistency between RQ1 and RQ3 strengthens both findings. 

Low-quality code risks being propagated with highly-voted, expert-authored, or high-performing notebooks acting as pedagogical templates. 
None of the studied variables, that a newcomer may naturally trust, offers any guarantee about the quality of a notebook.
Popularity is decoupled from quality (RQ1), and code expertise is decoupled
from both quality and performance (RQ3). 
Only the competition expertise gives a medium indication on the performance and ML quality but not on general Python code quality.


\subsection{Most common violations: our dataset violations against the literature}
\label{sec:discussion:literature}

\begin{table}[htbp]
\centering
\caption{ML specific SonarQube rules violation by mean density and absolut count (ordered by mean density)}
\label{tab:sonar_violations}
\begin{tabularx}{\textwidth}{c X c c c}
\toprule
\# & \textbf{Rule} & \textbf{Mean} & \textbf{Violation count} & \% \\
\midrule
1 & \texttt{dtype} parameter should be provided when using \texttt{pandas.read\_csv} or \texttt{pandas.read\_table} (S6740 - MAJOR) & 2.23e-02 & 765 400 & 87.25 \\
2 & Results that depend on random number generation should be reproducible (S6709 - MAJOR) & 2.68e-03 & 101 403 & 19.65 \\
3 & Important hyperparameters should be specified for machine learning libraries' estimators and optimizers (S6973 - MAJOR) & 2.41e-03 & 95 908 & 17.97 \\
4 & inplace=True should not be used when modifying a Pandas DataFrame (S6734 - CRITICAL) & 2.10e-03 & 87 013 & 12.18 \\
5 & When using pandas.merge or pandas.join, the parameters on, how and validate should be provided (S6735 - MAJOR) & 1.86e-03 & 111 181 & 13.25 \\
6 & numpy.random.Generator should be preferred to numpy.random.RandomState (S6711 - MAJOR) & 8.28e-04 & 65 466 & 8.80 \\
7 & \texttt{memory} parameter should be specified for Scikit-Learn Pipeline (S6969 - MINOR) & 7.88e-04 & 41 735 & 6.05 \\
8 & The \texttt{pandas.DataFrame.to\_numpy()} method should be preferred to the \texttt{pandas.DataFrame.values} attribute (S6741 - MAJOR) & 6.18e-04 & 27 661 & 4.95 \\
9 & np.nonzero should be preferred over np.where when only the condition parameter is set (S6729 - CRITICAL) & 4.97e-04 & 33 830 & 6.32 \\
10 & The \texttt{num\_workers} parameter should be specified for \texttt{torch.utils.data.DataLoader} (S6983 - MINOR) & 4.65e-04 & 29 079 & 4.70 \\
11 & Deprecated NumPy aliases of built-in types should not be used (S6730 - MAJOR) & 2.20e-04 & 12 748 & 2.64 \\
12 & Floating point numbers should not be tested for equality (S1244 - MAJOR) & 1.55e-04 & 11 219 & 1.74 \\
13 & The reduction axis/dimension should be specified when using reduction operations (S6929 - MAJOR) & 1.28e-04 & 9 544 & 1.77 \\
14 & \texttt{torch.tensor} should be used instead of \texttt{torch.autograd.Variable} (S6979 - MAJOR) & 4.74e-05 & 3 242 & 0.30 \\
15 & pandas.pipe method should be preferred over long chains of instructions (S6742 - MAJOR) & 2.63e-05 & 1 056 & 0.32 \\
16 & Subclasses of Scikit-Learn's \texttt{BaseEstimator} should not set attributes ending with \texttt{\_} in the \texttt{\_\_init\_\_} method (S6974 - CRITICAL) & 1.56e-05 & 1 541 & 0.26 \\
17 & \texttt{tf.function} should not depend on global or free Python variables (S6911 - MAJOR) & 1.35e-05 & 1 280 & 0.14 \\
18 & \texttt{model.eval()} or \texttt{model.train()} should be called after loading the state of a PyTorch model (S6982 - MAJOR) & 1.13e-05 & 1 441 & 0.28 \\
19 & Equality checks should not be made against \texttt{numpy.nan} (S6725 - BLOCKER) & 5.09e-06 & 333 & 0.07 \\
20 & Python side effects should not be used inside a \texttt{tf.function} (S6928 - CRITICAL) & 2.16e-06 & 197 & 0.04 \\
21 & Subclasses of \texttt{torch.nn.Module} should call the initializer (S6978 - MAJOR) & 1.93e-06 & 124 & 0.05 \\
22 & Nested estimator parameters modification in a Pipeline should refer to valid parameters (S6972 - MAJOR) & 3.71e-07 & 12 & 0.005 \\
23 & Einops pattern should be valid (S6984 - CRITICAL) & 2.43e-08 & 4 & 0.0008 \\
24 & PostgreSQL database passwords should not be disclosed (S6698 - BLOCKER) & 2.36e-08 & 1 & 0.0004 \\
25 & Transformers should not be accessed directly when a Scikit-Learn Pipeline uses caching (S6971 - CRITICAL) & 1.81e-08 & 3 & 0.001 \\
\bottomrule
\end{tabularx}
\end{table}

Our dataset allows us to easily observe the distribution of the Pylint violations and to compare it with the literature. 
The distribution of SonarQube violations is not compared against related work since, to our knowledge, it has not been studied in the literature.
Table~\ref{tab:sonar_violations} reports the SonarQube highest mean violation density (number of violations per SLOC) across our dataset with the absolute number of violations and the percentage of studied notebooks in which they occur. 
Table~\ref{sec:appendix:pylint_violations} reports, for each Pylint severity, the rules with
the highest mean violation density (number of violations per SLOC) across our dataset, together with the rate of studied notebooks in which they occur. 

\paragraph{Our most violated ML rules}
Table~\ref{tab:sonar_violations} reports the 25 ML rules violated at least once
in our dataset, ordered by mean violation density. 
The distribution is heavily skewed. 
A single rule, \emph{dtype parameter should be provided when using pandas.read\_csv or pandas.read\_table} (S6740), dominates the corpus: its mean density is an order of magnitude higher than that of the second-ranked rule,
and it occurs in 87.25\% of the studied notebooks. 
This rule is triggered when data is loaded through the pandas read functions without specifying column types, forcing the library to infer a type for every column:

\begin{verbatim}
# Noncompliant
pd.read_csv("my_file.csv")                                   
# Compliant
pd.read_csv("my_file.csv", dtype={'name': 'str', 'age': 'int'}) 
\end{verbatim}

Type inference is costly on large files and can silently assign unintended
types. 
Its large presence indicates that practitioners systematically delegate typing to pandas, regardless of these risks. 
More generally, three of the five most violated rules target the pandas API (S6740, S6734, S6735, S6741). 
This concentration is unsurprising, pandas is one of the most used data-handling librarie in the ML Python ecosystem~\cite{raschka2020a}.
Pandas-related rules apply to most of the notebooks, whereas framework-specific rules only apply to the subset of notebooks using the corresponding library.

Beyond data manipulation, two of the most violated rules concern the reproducibility of random number generation (S6709, S6711).
Random number generation is highly leveraged in ML pipelines, where it is used to randomly split training and evaluation data and for model training.
Fixing the randomness makes the results reproducible.

At the other end of the distribution, the least violated rules target narrow
misuses of specific frameworks (TensorFlow, PyTorch, einops). 
Nine rules are never violated (S6714, S6727, S6894, S6900, S6908, S6918, S6919, S6925,
S6985), and some are violated only anecdotally. 
Rule S6698 is triggered by a single notebook, due to a commented-out PostgreSQL connection string containing an identifier. 
Two factors explain these low counts. 
First, as noted above, a rule can only be violated by notebooks importing the corresponding library, so raw densities confound adherence to a practice with the prevalence of the library itself. 
Second, these rules capture fine-grained API misuses that are less likely to occur than the omission of an optional parameter.

Finally, prevalence does not imply impact: the most frequently violated rules are not necessarily those most strongly associated with lower performance. 
Our study measures the aggregate association between ML violations and performance (RQ2).
Quantifying the individual contribution of each rule is left as future work (Section~\ref{sec:discussion:researchers}).

\paragraph{The selected Pylint rules.} As detailed in \ref{sec:quality_rules:pylint} our analysis is purely static and does not restore the execution environment of the notebooks. 
As a consequence, our Pylint configuration does not resolve imports, and the import-resolution family of checks (e.g import-error E0401, no-member E1101, no-name-in-module E0611) is absent from our results. 
This methodological choice explains most of the divergences with the existing studies.
This family of checks dominates the Error category in previous ML studies.
Our methodology does not cause us to lose too much information, since these same studies repeatedly describe these checks as largely artificial.
Indeed, in \cite{simm20a}, the authors explicitly note that the \texttt{import-error} ``should be ignored'' as an artefact of not installing project dependencies.
The study in \cite{van21a} shows that \texttt{no-member} is massively false-positive on libraries written in C, such as PyTorch and other popular ML libraries.
Finally, in \cite{siddik2023a}, the authors recommend restoring the execution environment before drawing conclusions from such checks because of the high rate of false positives otherwise. 
Another difference is the tool version. 
Pylint's rule set changes over time; checks present in one study may be renamed or removed in another. 
We ran Pylint~\texttt{4.02}, which removed checks that older studies rely on (e.g. the
whitespace checks C0326 that \cite{quaranta2022a} report as their single
most frequent rule) and added new checks (e.g. \texttt{too-many-positional-arguments} R0917, \texttt{consider-using-f-string} C0209).

\paragraph{The way to report violations.} The studies we compare against do not all report violations the same way. 
\cite{adams2023a} report the percentage of files containing a given issue.
\cite{quaranta2022a} and \cite{van21a} report absolute counts.
\cite{simm20a} report the violations per non-blank line of code. 
We report both the mean density per SLOC and the share of notebooks with at least one occurrence in percent. 
These two axes for reporting violations can rank rules very differently. 
Some rules are violated only once per file, in our data \texttt{missing-module-docstring} (C0114) is present in \(99.39\%\) of notebooks yet has a near-zero density $0.0078$, whereas \texttt{line-too-long} (C0301) ranks high on both axes ($98.29\%$, density $0.090$). 
In the following we compare our findings regarding the number of violations in our dataset with the existing literature.
We used the percentage of files containing a given issue to compare ourselves with 
\cite{adams2023a}, and the
violation density with \cite{quaranta2022a},\cite{simm20a} and \cite{van21a}.

\paragraph{Siddik et al.~\cite{siddik2023a}.}
The most direct point of comparison is the work of \cite{siddik2023a}, who studied the same data source (Kaggle), reported Pylint issues by the same severities, and focused on ML notebooks.  
In their study, the authors report the 5 most violated Pylint rules (in absolute frequency) per severity.
For the Convention severity, our five most violated rules \texttt{line-too-long}, \texttt{trailing-whitespace}, \texttt{wrong-import-position}, \texttt{invalid-name}, and
\texttt{missing-function-docstring} are exactly the set they report among
their top issues.
For the  Convention violations, the only difference lies in the ordering, which can be explained by the fact that they ranked the rules according to the absolute number of violations, whereas we rank them based on violation density.

For the Warning rules, four of our top five (\texttt{unused-import}, \texttt{redefined-outer-name}, \texttt{bad-indentation}, \texttt{reimported}) coincide with theirs. The only rule they report and we do not, \texttt{pointless-statement} (W0104), is one we disabled as a notebook false positive (see Section~\ref{sec:quality_rules:pylint}). 
The authors stated that they used the default configuration for Pylint, even if some violations can be seen as false positives like import errors, because those errors are "what developers and users will run into when using notebooks from others"~\cite{siddik2023a}. 
While we agree on their take for import errors, the \texttt{pointless-statement} violation is mostly due to the medium (i.e, the notebook) itself and can be ignored.

For the error violations, the divergence is explained by import resolution.
Once their import-dependent checks (E0401, E1101, E0611) are set aside, we agree on \texttt{undefined-variable} (E0602) and \texttt{too-many-function-args} (E1121), which are our first and fifth most violated error rules, respectively. 

Both datasets share a highly similar set of the most frequently violated Pylint rules. 
Since both studies rely on the same underlying data source, this consistency reinforces the reliability of the violation detection across both works.

\paragraph{Adams et al.~\cite{adams2023a}.}
In their work, Adams et al.~\cite{adams2023a} reproduce the work of \cite{grotov2022a} on Kaggle ML notebooks and scripts.
\cite{grotov2022a} compare quality of notebooks and scripts on a large GitHub corpus (not restricted to ML) using Hyperstyle, which aggregates the Pylint, flake8, and WPS quality tools.  
Regarding the Pylint violations they found in their dataset, the most violated rules in the Warning categorie are \texttt{unused-import} (W0611), \texttt{redefined-outer-name} (W0621), \texttt{reimported} (W0404), \texttt{unused-variable} (W0612), \texttt{unused-argument} (W0613), \texttt{unnecessary-semicolon} (W0301), \texttt{bad-indentation} (W0311) and \texttt{pointless-statement} (W0104). 
We share those violations as our most violated Warning rules; the only exception is the \texttt{pointless-statement}, the one we disable as a notebook false positive (see Section~\ref{sec:quality_rules:pylint}).
For the Convention violations, they report \texttt{wrong-import-order} (C0411), \texttt{ungrouped-imports} (C0412) and \texttt{trailing-newlines} C0305. 
Respectively, our 6, 7, and 17 most violated Convention rules.
For the Error, they report \texttt{no-member} (E1101), \texttt{syntax-error} (E0001), and \texttt{no-name-in-module} E0611. 
We share none of them, as the \texttt{no-member} and \texttt{no-name-in-module} are dependency-aware rules that we deactivated (see Section~\ref{sec:quality_rules:pylint}), and the notebooks containing \texttt{syntax-error} were dropped from our dataset due to risk of biasing the results (see Section~\ref{para:nb_convertion:pylint_violation}).

Both observations share similarities; we have the same most violated Warning rules, but our top Convention violation seems to differ, and we have no clear explanation why. 
This may come from the tool they used (Hyperstyle) or from the new violation categories they defined (Best Practices, Code Style, and Error Proneness), since they only share the 5 most violated rules in each category and maybe not all the Convention rules belong to their categories.
We can not compare our Error and Refactor violations due to the configuration difference and the absence of Refactor violations in their results. 


\paragraph{Quaranta et al.~\cite{quaranta2022a}.}
In their work, Quaranta et al.~\cite{quaranta2022a} also lint Kaggle notebooks with Pylint, on a curated set
of $1,380$ collaboratively authored, expert, medal-awarded notebooks. 
Their reported top violation \texttt{bad-whitespace} no longer exists in our
Pylint version, having been removed from Pylint in release 2.6.0.
Setting the removed checks aside, their ranking (\texttt{invalid-name}, \texttt{trailing-whitespace}, \texttt{redefined-outer-name}, \texttt{unused-import}, \texttt{undefined-variable}, \texttt{too-many-locals}, \texttt{too-many-arguments}, \texttt{no-else-return}) substantially overlaps ours within the corresponding severities. 

However, we do \emph{not} reproduce their upvote effect.
Quaranta et al.~\cite{quaranta2022a} report that the most upvoted notebooks violate fewer Pylint rules (e.g., the share violating warning rules falls from $68.39\%$ overall to $45\%$ in their top decile).
The divergence may be explained by differences in design. 
Their dataset is pre-filtered to expert, collaboratively authored, medal-winning notebooks and the observed effect is small.
A limitation that the authors pointed out: "despite a slight tendency for the most upvoted notebooks to be more compliant with the best practices, the small differences observed in the three samples are arguably explained by the fact that the original dataset in our study already comprises high-quality notebooks, as they are filtered according to authors’ expertise".
Our sampling criteria admit a wider quality range that may erase such effects.


\paragraph{Oort et al.~\cite{van21a}.}
In their work, Oort et al.~\cite{van21a} studied ML projects (74 ML projects including 32 projects from finished Kaggle competitions, 38 from paperswithcode.com, and 4 from reproducedpapers) by restoring each project's environment before linting using Pylint.
Their Error category is therefore led by \texttt{no-member} and \texttt{import-error}, and their top refactoring smell is \texttt{duplicate-code}.
Precisely the dimensions our configuration cannot assess, since we neither resolve imports nor enable duplication detection (in our configuration, we linted one file at a time, which disables cross-file duplicate detection).
The rules on which we most visibly diverge from Oort et al.~\cite{van21a} are exactly those that our methodology excludes.

Setting aside those excluded dimensions, the overlap between the two studies is substantial across all four categories.
For Convention, fifteen of their top 20 violation appear directly in ours.
The remaining variations are attributable to Pylint version changes or may come from the difference in the project selection.
They report (\texttt{bad-whitespace} (C0326), \texttt{missing-docstring} (C0111) that were removed or renamed in newer versions of Pylint, and some violations that appear in our results (\texttt{consider-using-f-string} (C0209) and \texttt{unnecessary-lambda-assignment} (C3001)) were introduced in later releases.
For Refactor, beyond \texttt{duplicate-code}, twelve of their rules appear in our top 20. 
For Warning, their ranking is dominated by \texttt{unused-wildcard-import}, which the authors attribute to a single outlier project responsible for over half of those violations. 
Outside this artefact, 10 of their warning violations recur in ours.
For Error, once the three dependency-aware rules (\texttt{no-member}, \texttt{import-error}, \texttt{no-name-in-module}) are set aside, both studies identify the same dominant error: \texttt{undefined-variable} (E0602). 
The \texttt{bad-option-value} no longer exists in our Pylint version.
Agreement is found on 11 of the remaining Error violations.

Our most violated rules seem to be close to the ones reported by Oort et al.~\cite{van21a}. 
Some divergences are attributable to Pylint version changes and methodological differences in import resolution or duplication detection.
The other divergence might be explained by the difference in project selection and the sample size.

\paragraph{Summary.}
In the literature, we found no prior work studying the violations of SonarQube ML rules.
A line of work exists on the verification of ML-specific rules~\cite{haakman2020a,van2022a,shivashankar2025a,dolcetti2026a}, but none of them are directly comparable to our selected ML rules.
Regarding the Pylint Python violations, the comparison supports the following two conclusions. 
First, the Pylint violations of our dataset are generally consistent with prior work once tool configuration is fixed (import resolution, differences in the rule set and, clone detection). 
The residual differences might be attributable to sampling strategies and sample size.
Second, the negligible correlation between popularity (upvote) and general Python code quality (Pylint violation) is not consistent with the work of Quaranta et al.~\cite{quaranta2022a}. 
This difference might be explained by the notebook sampling and the sample size, as hypothesized by Quaranta et al.

\subsection{Implications}
\label{sec:discussion:implications}
 
\subsubsection{For ML newcomers}
\label{sec:discussion:newcomers}
None of the variables we observed on Kaggle reliably indicates whether a notebook contains high-quality code (SE-practices) or not. 
Upvotes do not (RQ1), and neither does an author's code-expert status (RQ3). 
Newcomers who select notebooks by popularity or by author badge may inherit poor SE-practices.
Only the competition expertise can give a small hint on the respect of good ML-practices. 
Recommendation is to learn software engineering practices alongside machine learning rather than to infer them from existing notebooks and to treat cloned notebooks as starting points to be reviewed, not as validated examples.
 
\subsubsection{For ML practitioners}
\label{sec:discussion:practitioners}
For ML-specific practices, our observations suggest a tangible upside: fewer ML violations are associated with better performance, consistently and across contexts.
Adopting these practices early is, at worst, neutral and plausibly beneficial to outcomes. 
As noted in Section~\ref{sec:discussion:constructs}, for general Python (SE) quality, we found no performance effect in either direction.
Applying Pylint conventions is not correlated with better or worse performance; this means that applying Pylint conventions costs nothing in performance. 
No performance gain does not mean no benefit, the gains of clean code lie in reproducibility, reuse, and comprehension.
Dimensions our study does not capture. 
The combination of zero measured performance cost and well-established maintainability benefits makes adopting best SE-practices from the earliest stages of development a rational default rather than a costly afterthought.
 
\subsubsection{For researchers}
\label{sec:discussion:researchers}
Our findings motivate two directions for researchers on the topic. 
First, the ML-specific rule set should be expanded and characterised: with only 34 rules yielding small effects, defining additional rules and quantifying each rule's individual association with performance would both sharpen the effect estimate and identify which practices matter most. 
Defining more rules may also increase the observed effects.
Second, because clone-and-own could be one of the vectors through which bad practices propagate, violations and code clones should be studied jointly.
Understanding which low-quality patterns are most frequently reused would let the community target remediation where its ecosystem-wide impact is greatest. Similarly, investigating the distribution of violations across the different steps of the ML pipeline~\cite{biswas2022a} seems to be an interesting direction. This would allow us to observe whether certain violations predominantly occur during specific steps, and to assess the individual impact of these step-specific violations on model performance.

Researchers using Kaggle as a data source should be warned that, following our results and unlike Stack Overflow where snippets with more upvotes contain higher quality code\cite{bafatakis2019a}, upvotes and expertise do not necessarily mean high quality or better performance. This information needs to be taken into consideration when designing sampling methodologies.

\section{Limitations and Threats to Validity}
\label{sec:threats}
We discuss the threats to the validity of our study and the choices made to mitigate them, organised along the four categories: construct, internal, conclusion, and external validity.

\subsection{Construct Validity}
\label{sec:threats-construct}

Our notebook selection pipeline does not account for the date at which a notebook was authored. 
The oldest notebooks in our dataset were authored in February 2017 and the most recent in January 2026. 
The oldest notebooks likely rely on outdated practices and library versions. 
For Pylint, we manually set a minimum Python version (see Section~\ref{filter_3_tools}), so version-dependent rules should not be triggered by outdated language constructs. 
For the SonarQube ML rules, however, we did not set any version threshold, and these rules were
designed for recent ML APIs and libraries.
Our SonarQube ML rules may therefore be less accurate on older code and risk yielding false positives. 
We consider this threat limited.
Practitioners on Kaggle reuse and learn from older notebooks regardless of their authoring date, so measuring recent best practices against this corpus reflects the bad-practices our study investigates.

We use the Kaggle competition score as a proxy for a notebook's ML performance.
This choice has limits; we cannot guarantee that every notebook submitted to a competition genuinely attempts to maximise that score. 
Some submissions may not be true competition entries and may not be intended to
perform well, which could bias our observations. 
To mitigate this, the \textit{Best-Per-Team Dataset} retains only the highest-performing notebook per team and per competition (see Section~\ref{best_per_team_dataset}).
We also applied a minimum threshold of 30 source lines of code to exclude notebooks that do not build ML pipeline (discussed further in \ref{sec:threats-internal}).

The choice of static analysis tools is itself a threat. 
We selected Pylint and SonarQube (see Section~\ref{sec:code_quality_tool_choice}), but a different tool set (and by extension, a different rule set) could yield different effects and effect sizes. 
More broadly, static analysis tools are known to produce false positives~\cite{guo2023a}, which may add
noise to our quality measurements.

The weights assigned to violation severities in both quality scores could threaten construct validity. 
For the Pylint quality score, we kept the original formula unchanged. 
Across all our tests, the Pylint violation count and the Pylint quality score point in the same direction, with only small differences in effect size. 
These differences in effect sizes are too small for any effect to rise above the negligible range. 
For the SonarQube ML quality score, we reused the Pylint formula and weighted only the Blocker severity ($\times 5$).
The Blocker level is the most severe level, analogously to Pylint errors. 
This choice does not affect our conclusions: the heavily weighted Blocker
severity is almost absent from the corpus (only $334$ violations, see Table~\ref{tab:sonar_violations}).

On the user's expertise level, we made two decisions: 
First, we collapsed the five Kaggle tiers (novice, contributor, expert, master and grand master) into two groups, expert and non-expert (see Section~\ref{users_expertise}).
This loses the fine granularity of the five-tier system but preserves the distinction between users who have received community or competitive validation and those who have not. 
Second, for team authored notebooks, we retained the expertise level of the most experienced member. 
This could bias our analysis, since a single member's tier is used to characterise the whole
team. 
We consider this bias acceptable because most notebooks are single authored (discussed in Section~\ref{users_expertise}) and, more importantly, the cases where a team shares non-expert and expert users are rare.
Only $2.5\%$ of the whole dataset are team-authored notebooks that mix code experts and non-experts ($2.69\%$ have non-uniform code expertise), and
only $1.07\%$ mix competition experts and non-experts ($1.92\%$ have
non-uniform competition expertise). 
Given these small proportions, the aggregation choice is unlikely to affect our results.

We standardise the private competition score with a within-group z-score (Section~\ref{z-score}) to overcome the fact that raw scores are tied to a specific competition and are not comparable across competitions (different tasks, metrics, ranges, and polarities). 
The z-score expresses a notebook's relative standing within its own competition rather than an
absolute level of performance, and it assumes a meaningful group means and standard deviation. 
It is therefore less reliable for competitions with few submissions or with strongly skewed score distributions. 
Notebooks for which a z-score could not be computed were excluded. 
We consider this standardisation sufficient for comparing relative performance across heterogeneous competitions, but it does not support conclusion about absolute performance.

\subsection{Internal Validity}
\label{sec:threats-internal}

Teams frequently iterate by cloning, modifying, and resubmitting a notebook, producing several near-identical observations while following our protocol.
Treating each submission as independent would lead to pseudo-replication, over-representing the coding habits of highly active teams. 
We neutralise this intra-team dependency by construction, keeping a single notebook per team and per competition in the \textit{Best-Per-Team Dataset} (discussed in Section~\ref{coufounding_variables}), so that observations are independent.

Two further structural biases cannot be removed by construction, so we assessed
them empirically (Section~\ref{coufounding_methodo}). 
The selection bias, whether team-selected notebooks differ from non-selected drafts, is small regarding the performance and negligible regarding the number of violations. 

The post-deadline submission bias (whether notebooks submitted after the deadline
benefit from knowledge spillover, and whether the absence of a time constraint
affects quality) is negligible across performance, Pylint violations, and ML
violations (Table~\ref{tab:mann_whitney_coufunding_deadline}). 
The latter result also indicates that the competition time constraint (when practitioners may priorise ML performance over code quality because of the short time period imposed by the competition) does not noticeably affect notebook quality. 
We took these results into account when defining the analysis samples (Section~\ref{final_sampling}).

During notebook selection, we excluded notebooks with fewer than 30 source lines of code to remove those unlikely to contain ML pipeline code. 
Such notebooks would bias the study because they carry a performance score and quality scores, while the code they contain does not produce that performance (e.g. notebooks uploading pre-computed results, or trivial code). 
Even with this threshold, some remaining notebooks may contain code that is not directly responsible for the performance output.
In those cases, the code we analyse is not the code that produced the score, which introduces bias. 
This threat is the counterpart, on the internal side, of the performance proxy limitation discussed previously
in Section~\ref{sec:threats-construct}.

\subsection{Conclusion Validity}

We exclusively used non-parametric tests (Spearman's rank correlation and the Wilcoxon Mann Whitney U test) to avoid restrictive assumptions about the distribution of our data. 
This ensures robustness in the presence of non-normally distributed data and outliers.
For every test, we conducted an \textit{a priori} sensitivity analysis (power $0.95$ for the research questions, $0.99$ for the confounder assessments, $\alpha = 0.005$) and did not interpret any effect below the corresponding minimal detectable threshold. 
For the same reason, we excluded the competition types whose sample sizes were too small to detect even small effects at our target power (Community, Recruitment, and Analytics. Section~\ref{sec:rq2_methodo:competition_type}).

We report a large number of statistical tests and deliberately do not apply a multiple comparison correction to our p-values. 
We consider this acceptable because our conclusions rely on the agreement of many tests, and $p$-values span many orders of magnitude below $alpha$.
Moreover, our significance threshold $\alpha = 0.005$ is already conservative compared to the $\alpha = 0.05$ standard.

\subsection{External Validity}

We studied ML code extracted from Jupyter notebooks. 
Our findings may not hold for code written outside the notebook medium.
Prior work reports that code quality in Python notebooks is lower than in scripts~\cite{grotov2022a}, suggesting that the medium itself may influence quality. 

All analysed code was extracted from notebooks submitted to Kaggle competitions. 
Kaggle practitioners may form a specific population, and our findings may not generalise to other communities. 
They may not even generalise to Kaggle as a whole, since we studied the subset of users participating in competitions and only code authored in a competition context.

We restricted our analysis to Python, the most widely used language on Kaggle and in ML. 
Our findings offer no guarantee for notebooks written in other languages, such as R.

Our results are based on a single Kaggle data snapshot (January 2026).
As practices, libraries, and the platform's tier system evolve, the observed relationships may shift over time.

\section{Conclusion}
\label{sec:conclusion}
In this paper, we conducted a large-scale empirical study of 265,363 Python machine learning notebooks submitted to Kaggle competitions. 
We assessed the quality of each notebook with two complementary static analysis tools: Pylint, capturing general Python code quality, and SonarQube, configured with a profile of 34 rules targeting data-science and ML-specific practices. 
We then related rule violations to three dimensions of the Kaggle ecosystem: ML performance (measured by the private competition score, standardized within competitions), popularity (upvotes), and author expertise (Kaggle progression tiers).

Our results show that the widespread assumption that code quality and ML performance are independent does not hold uniformly.
The relation between code quality and performance depends on which notion of quality is considered. 
General Python code quality is effectively decoupled from ML performance: across all our tests, Pylint violations show negligible or non-significant correlations with competition scores. 
In contrast, ML-specific violations exhibit a consistent small negative correlation with performance, an association that persists when stratifying by competition type and by author expertise. 
In other words, following general software engineering practices costs nothing in performance, while following ML-specific practices is associated with better outcomes. 

Regarding the social signals available on the platform, neither popularity (RQ1) nor code expertise (RQ3) provides any indication of quality or performance.
Only competition expertise relates to better performance and fewer ML-specific violations, without any guarantee concerning the general Python quality.

These findings carry implications for the ML community. 
Since high-quality code is compatible with high performance (ML-specific practices possibly conducive to better performance), software engineering best practices can be integrated from the earliest stages of experimentation rather than treated as a costly afterthought. 
Newcomers and researchers should note that highly-voted or expert-authored notebooks offer no guarantee of quality, and sampling strategies relying on these signals should be reconsidered.
As future work, we plan to expand and characterize the ML-specific rule set to identify which individual practices matter most for performance, to study jointly code clones and violations to understand how poor practices propagate through the clone-and-own culture, and to locate violations within the different steps of the ML pipeline.



\section*{Declarations}
\label{sec:declaration}

\noindent\textbf{Author Contributions} All authors contributed to the conception and design of the study. 
All authors contributed to the writing, the first author for the original draft preparation and the other authors to review and editing.
The first author conducted the data collection (i.e., ran the experiments). 
The first author performed most of the data analysis. When uncertainties arose, all authors discussed and contributed to the interpretation of the results.
\newline

\noindent\textbf{Funding} This work was supported by the French National Research Agency (ANR) under project ANR-AAPG2024, PROFIL.
\newline

\noindent\textbf{Data availability statement} The datasets generated and analysed during the current study and the R/Python analysis scripts are available in the following repository, \url{https://doi.org/10.5281/zenodo.21464700}. 
\newline

\noindent\textbf{Ethical approval} Not applicable.
\newline

\noindent\textbf{Informed consent} Not applicable.
\newline

\noindent\textbf{Conflicts of interests/Competing interests:} The authors declare that they have no known competing financial interests or personal relationships that could have appeared to influence the work reported in this paper.

%
%

\bibliographystyle{spmpsci}      
\bibliography{references}   

%
%

\appendix

\section{SonarQube Selected Rules}
\label{detail_sonar_selected_rules}
Our SonarQube analysis profile comprise the following 34 python rules displayed in Table~\ref{tab:sonarqube_rules}.

\begin{longtable}{
    p{\dimexpr 0.10\textwidth - 2\tabcolsep} 
    p{\dimexpr 0.35\textwidth - 2\tabcolsep} 
    p{\dimexpr 0.55\textwidth - 2\tabcolsep}
}
\caption{SonarQube Python Rules Profile} \label{tab:sonarqube_rules} \\
\toprule
\textbf{Code} & \textbf{Title} & \textbf{Description} \\
\midrule
\endhead 

\bottomrule
\endfoot

S1244 & Floating point numbers should not be tested for equality & This rule raises an issue when direct and indirect equality/inequality checks are made on floats. Floating point math is imprecise because of the challenges of storing such values in a binary representation. Therefore, the use of the equality (==) and inequality (!=) operators on float values is almost always erroneous. \\ \midrule

S6709 & Results that depend on random number generation should be reproducible & This rule raises an issue when random number generators do not specify a seed parameter. To ensure that results are reproducible, it is important to use a predictable seed in this context. \\ \midrule

S6711 & numpy.random.Generator should be preferred to numpy.random.RandomState & This rule raises an issue when legacy numpy.random.RandomState is used. The preferred best practice to generate reproducible pseudorandom numbers is to instantiate a numpy.random.Generator object with a seed and reuse it in different parts of the code. This avoids the reliance on a global state. Whenever a new seed is needed, a new generator may be created instead of mutating a global state. \\ \midrule

S6714 & Passing a list to np.array should be preferred over passing a generator & This rule raises an issue when a generator is passed to np.array. \\ \midrule

S6725 & Equality checks should not be made against "numpy.nan" & This rule raises an issue when an equality check is made against numpy.nan. Equality checks of variables against numpy.nan in NumPy will always be False due to the special nature of numpy.nan. This can lead to unexpected and incorrect results. Instead of standard comparison the numpy.isnan() function should be used. \\ \midrule

S6727 & The abs\_tol parameter should be provided when using math.isclose to compare values to 0 & This rule raises an issue when math.isclose is used to compare values against 0 without providing the abs\_tol parameter. Comparing float values for equality directly is not reliable and should be avoided, due to the inherent imprecision in the binary representation of floating point numbers. When using math.isclose, the absolute tolerance is defined through the parameter abs\_tol. By default, the value of this parameter is 0.0. Therefore, using math.isclose to compare values against zero without providing this parameter is equivalent to a strict equality check, which is likely not intended. \\ \midrule

S6729 & np.nonzero should be preferred over np.where when only the condition parameter is set & This rule raises an issue when np.where is used with only the condition parameter set. When providing only the condition parameter to the np.where function, it will behave as np.asarray(condition).nonzero() or np.nonzero(condition). Both these functions provide a way to find the indices of the elements satisfying the condition passed as parameter. Be mindful that np.asarray(condition).nonzero() and np.nonzero(condition) do not return the values that satisfy the condition but only their indices. This means the result variable now holds a tuple with the first element being an array of all the indices where the condition arr $>$ 2 was satisfied: (array([2,3]),). If the intention is to find the indices of the elements which satisfy a certain condition it is preferable to use the np.asarray(condition).nonzero() or np.nonzero(condition) function instead. \\ \midrule

S6730 & Deprecated NumPy aliases of built-in types should not be used & This rule raises an issue when a deprecated Numpy alias of a built-in type is used. \\ \midrule

S6734 & "inplace=True" should not be used when modifying a Pandas DataFrame & This rule raises an issue when the inplace parameter is set to True when modifying a Pandas DataFrame. Using inplace=True when modifying a Pandas DataFrame means that the method will modify the DataFrame in place, rather than returning a new object. Generally speaking, the motivation for modifying an object in place is to improve efficiency by avoiding the creation of a copy of the original object. Unfortunately, many methods supporting the inplace keyword either cannot actually be done inplace, or make a copy as a consequence of the operations they perform, regardless of whether inplace is True or not. Additionally, using inplace=True may trigger a SettingWithCopyWarning and make the overall intention of the code unclear. \\ \midrule

S6735 & Required parameters should be used for pandas.merge or pandas.join & This rule raises an issue when the parameters how, on and validate are not provided when using pandas.merge or pandas.join. \\ \midrule

S6740 & Required parameters should be used for pandas.read\_csv or pandas.read\_table & This rule raises an error when the dtype parameter is not provided when using pandas.read\_csv or pandas.read\_table. \\ \midrule

S6741 & The pandas.DataFrame.to\_numpy() method should be preferred to the pandas.DataFrame.values attribute & This rule raises an issue when the pandas.DataFrame.values is used instead of the pandas.DataFrame.to\_numpy() method. \\ \midrule

S6742 & The pandas.pipe method should be preferred over long chains of instructions & This rule raises an issue when 7 or more commands are applied on a data frame. The pandas library provides many ways to filter, select, reshape and modify a data frame. Pandas supports as well method chaining, which means that many DataFrame methods return a modified DataFrame. This allows the user to chain multiple operations together, making it effortless perform several of them in one line of code. To improve code readability, debugging, and maintainability, it is recommended to break down long chains of pandas instructions into smaller, more modular steps. This can be done with the help of the pandas pipe method, which takes a function as a parameter. This function takes the data frame as a parameter, operates on it and returns it for further processing. Grouping complex transformations of a data frame inside a function with a meaningful name can further enhance the readability and maintainability of the code. \\ \midrule

S6894 & Dates should be formatted correctly when using "pandas.to\_datetime" with "dayfirst" or "yearfirst" arguments & This rule raises an issue when the argument dayfirst or yearfirst is set to True on pandas.to\_datetime function with an incorrect string format. The pandas.to\_datetime function transforms a string to a date object. The string representation of the date can take multiple formats. To correctly parse these strings, pandas.to\_datetime provides several arguments to setup the parsing, such as dayfirst or yearfirst. For example setting dayfirst to True indicates to pandas.to\_datetime that the date and time will be represented as a string with the shape day month year time. Similarly with yearfirst, the string should have the following shape year month day time. These two arguments are not strict, meaning if the shape of the string is not the one expected by pandas.to\_datetime, the function will not fail and try to figure out which part of the string is the day, month or year. \\ \midrule

S6900 & Numpy weekmask should have a valid value & This rule raises an issue when a numpy weekmask format is incorrect. To allow a datetime to be used in contexts where only certain days of the week are valid, NumPy includes a set of business day functions. Weekmask is used to customize valid business days. Setting an incorrect weekmask leads to ValueError. \\ \midrule

S6908 & "tensorflow.function" should not be recursive & This rule raises an issue when a \texttt{tensorflow.function} is recursive. When defining a \texttt{tensorflow.function} it is generally a bad practice to make this function recursive. TensorFlow does not support recursive \texttt{tensorflow.function} and will in the majority of cases throw an exception. However it is possible as well that the execution of such function succeeds, but with multiple tracings which has strong performance implications. When executing \texttt{tensorflow.function}, the code is split into two distinct stages. The first stage call tracing creates a new \texttt{tensorflow.Graph}, runs the Python code normally, but defers the execution of TensorFlow operations. These operations are added to the graph without being ran. The second stage which is much faster than the first, runs everything that was deferred previously. Depending on the input of the \texttt{tensorflow.function} the first stage may not be needed. Skipping this first stage is what provides the user with TensorFlow’s high performance. Having a recursive \texttt{tensorflow.function} prevents the user from benefiting of TensorFlow’s capabilities. \\ \midrule

S6911 & "tf.function" should not depend on global or free Python variables & This rule raises an issue when a tensorflow.function depends on a global or free Python variable. When calling a tensorflow.function behind the scenes a ConcreteFunction is created everytime a new value is passed as argument. This is not the case with Python global variables, closure or nonlocal variables. This means the state and the result of the tensorflow.function may not be what is expected. \\ \midrule

S6918 & "tf.Variable" objects should be singletons when created inside of a "tf.function" & This rule raises an issue when a tensorflow.Variable created inside of a tensorflow.function is not a singleton. tensorflow.functions only supports singleton tensorflow.Variables. This means the variable will be created on the first call of the tensorflow.function and will be reused across the subsequent calls. Creating a tensorflow.Variable that is not a singleton will raise a ValueError. \\ \midrule

S6919 & The "input\_shape" parameter should not be specified for "tf.keras.Model" subclasses & This rule raises an issue when the input\_shape is specified in a tensorflow.keras.Model subclass. Keras provides a full-featured model class called tensorflow.keras.Model. It inherits from tensorflow.keras.layers.Layer, so a Keras model can be used and nested in the same way as Keras layers. Keras models come with extra functionality that makes them easy to train, evaluate, load, save, and even train on multiple machines. As the tensorflow.keras.Model class inherits from the 'tensorflow.keras.layers' you do not need to specify input\_shape in a subclassed model; this argument will be ignored. \\ \midrule

S6925 & The "validate\_indices" argument should not be set for "tf.gather" function call & This rule raises an issue when the validate\_indices argument is set for `tf.gather ` function call. The tf.gather function allows you to gather slices from a tensor along a specified axis according to the indices provided. The validate\_indices argument is deprecated and setting its value has no effect. Indices are always validated on CPU and never validated on GPU. \\ \midrule

S6928 & Python side effects should not be used inside a "tf.function" & This rule raises an issue when a Python side effect happens inside a tensorflow.function. Python sides effects such as printing, mutating a list or a global variable, inside of a tensorflow.function may not behave as expected. Because of the Rules of tracing, the execution of side effects will depend on the input values of the function and will execute only once per tracing. \\ \midrule

S6929 & The axis argument should be specified when using TensorFlow's reduction operations & This rule raises an issue when the axis/dim` argument is not provided to reduction operations. \\ \midrule

S6969 & "memory" parameter should be specified for Scikit-Learn Pipeline & This rule raises an issue when a Scikit-Learn Pipeline is created without specifying the \textbf{memory} argument. When the \textbf{memory} argument is not specified, the pipeline will recompute the transformers every time the pipeline is fitted. This can be time-consuming if the transformers are expensive to compute or if the dataset is large. However, if the intent is to recompute the transformers everytime, the memory argument should be set explicitly to \textbf{None}. This way the intention is clear. \\ \midrule

S6971 & Transformers should not be accessed directly when a Scikit-Learn Pipeline uses caching & This rule raises an issue when trying to access a Scikit-Learn transformer used in a pipeline with caching directly. When using a pipeline with a cache and passing the transformer objects as an instance from a variable, it is possible to access the transformer objects directly. This is an issue since all the transformers are cloned when the Pipeline is fitted, and therefore, the objects outside the Pipeline are not updated and will yield unexpected results. \\ \midrule

S6972 & Nested estimator parameters adjustment in a Pipeline should refer to valid parameters & This rule raises an issue when an invalid nested estimator parameter is set on a Pipeline. In the sklearn library, when using the Pipeline class, it is possible to modify the parameters of the nested estimators. This modification can be done by using the Pipeline method set\_params and specifying the name of the estimator and the parameter to update separated by a double underscore \_\_. Providing invalid parameters that do not exist on the estimator can lead to unexpected behavior or runtime errors. This rule checks that the parameters provided to the set\_params method of a Pipeline instance or through the param\_grid parameters of a GridSearchCV are valid for the nested estimators. \\ \midrule

S6973 & Important hyperparameters should be specified for Scikit-Learn estimators & This rule raises an issue when a machine learning estimator or optimizer is instantiated without specifying the important hyperparameters. When instantiating an estimator or an optimizer, default values for any hyperparameters that are not specified will be used. Relying on the default values can lead to non-reproducible results across different versions of the library. Furthermore, the default values might not be the best choice for the specific problem at hand and can lead to suboptimal performance. \\ \midrule

S6974 & Subclasses of Scikit-Learn's "BaseEstimator" should not set attributes ending with "\_" in the "\_\_init\_\_" method & This rule raises an issue when an attribute ending with \_ is set in the \_\_init\_\_ method of a class inheriting from Scikit-Learn BaseEstimator. On a Scikit-Learn estimator, attributes that have a trailing underscore represent attributes that are estimated. These attributes have to be set in the fit method. Their presence is used to verify if an estimator has been fitted. \\ \midrule

S6978 & Subclasses of "torch.nn.Module" should call the initializer & This rule raises an issue when a class is a Pytorch module and does not call the super().\_\_init\_\_() method in its constructor. To provide the AutoGrad functionality, the Pytorch library needs to set up the necessary data structures in the base class. If the super().\_\_init\_\_() method is not called, the module will not be able to keep track of its parameters and other attributes. For example, when trying to instantiate a module like nn.Linear without calling the super().\_\_init\_\_() method, the instantiation will fail when it tries to register it as a submodule of the parent module. \\ \midrule

S6979 & "torch.tensor" should be used instead of "torch.autograd.Variable" & This rule raises when a torch.autograd.Variable is instantiated. The Pytorch Variable API has been deprecated. The behavior of Variables is now provided by the Pytorch tensors and can be controlled with the requires\_grad parameter. The Variable API now returns tensors anyway, so there should not be any breaking changes. \\ \midrule

S6982 & "model.eval()" or "model.train()" should be called after loading the state of a PyTorch model & This rule raises an issue when a PyTorch model state is loaded and \textbf{torch.nn.Module.eval()} or \textbf{torch.nn.Module.train()} is not called. When using PyTorch it is common practice to load and save a model’s state from/to a \textbf{.pth} file. Doing so allows, for example, to instantiate an untrained model and load learned parameters coming from another pre-trained model. Once the learned parameters are loaded to the model it is important, before inferencing, to clearly state the intention by calling \textbf{torch.nn.Module.eval()} method to set the model in evaluation mode or calling \texttt{torch.nn.Module.train()} to indicate the training will resume. Failing to call \textbf{torch.nn.Module.eval()} would leave the model in training mode which may not be the intention. \\ \midrule

S6983 & The "num\_workers" parameter should be specified for "torch.utils.data.DataLoader" & This rule raises an issue when a \texttt{torch.utils.data.Dataloader} is instantiated without specifying the num\_workers parameter. \\ \midrule

S6984 & Einops pattern should be valid & This rule raises an issue when an incorrect pattern is provided to an einops operation. The einops library provides a powerful and flexible way to manipulate tensors using the Einstein summation convention. The einops uses a different convention than the traditional one. In particular, the axis names can be more than one letter long and are separated by spaces. \\ \midrule

S6985 & Usage of "torch.load" can lead to untrusted code execution & This rule raises an issue when pytorch.load is used to load a model. In PyTorch, it is common to load serialized models using the torch.load function. Under the hood, torch.load uses the pickle library to load the model and the weights. If the model comes from an untrusted source, an attacker could inject a malicious payload which would be executed during the deserialization. \\

S6698 & PostgreSQL database passwords should not be disclosed & Secret leaks often occur when a sensitive piece of authentication data is stored with the source code of an application. Considering the source code is intended to be deployed across multiple assets, including source code repositories or application hosting servers, the secrets might get exposed to an unintended audience.

\end{longtable}

\section{Cohen's effect size conversion formula adapted for groups of different size}

\paragraph{User-selected notebooks group and non-selected notebooks group.}
\label{effect_size_convertion_selected_not_selected}
For the user-selected notebooks group ($n=11,918$) and non-selected notebooks group ($n=253,445$).
We found an effect $d=0.049$. 
Using the adapted Cohen's convertion formula: 

$$r = \frac{0.049}{\sqrt{0.049^2 + a}}$$

$$a = \frac{(11918 + 253445)^2}{11918*253445} $$

$$0.01 = \frac{0.049}{\sqrt{0.049^2 + 23.31}}$$

After convertion, we found an effect $r=0.01$.

\paragraph{Notebooks submitted before and after the competition deadline.}
\label{effect_size_convertion_before_after_deadline}
For the notebooks submitted before the competition deadline ($n=214,294$) and those submitted after ($n=51,069$).
We found an effect $d=0.026$.
Using the adapted Cohen's convertion formula: 

$$r = \frac{0.026}{\sqrt{0.026^2 + a}}$$

$$a = \frac{(214294 + 51069)^2}{214294*51069} $$

$$0.01 = \frac{0.026}{\sqrt{0.026^2 + 6.43}}$$

After convertion, we found an effect $r=0.01$.

\paragraph{RQ3. Competition expertise power analysis}
\label{rq3_annexe_convertion_effect}
$$r = \frac{0.0789314}{\sqrt{0.0789314^2 + a}}$$

$$a = \frac{(3620 + 41823)^2}{3620*41823} $$

$$0,021367 = \frac{0.0789314}{\sqrt{0.0789314^2 + 13.64}}$$

After convertion, we found an effect $r=0.021367$.

\paragraph{RQ3. Code expertise power analysis}
$$r = \frac{0.0609306}{\sqrt{0.0609306^2 + a}}$$

$$a = \frac{(6529 + 38914)^2}{6529*38914} $$

$$0,021364 = \frac{0.0609306}{\sqrt{0.06093064^2 + 8.13}}$$

After convertion, we found an effect $r=0.021364$.

\section{Power Analysis Parameters and Results Tables}
\label{g_power_appendix}

\begin{table}[h]
\centering
\caption{Sensitivity Analysis Parameters and Results (G*Power) on users selected vs not-selected notebooks}
\label{tab:gpower_sensitivity_wilcoxon_selected}
\begin{tabular}{ll}
\textbf{Parameter} & \textbf{Value} \\ \hline
\multicolumn{2}{l}{\textit{Test Configuration}} \\
Test family & $t$ tests \\
Statistical test & Means: Wilcoxon-Mann-Whitney test (two groups) \\
Options & A.R.E. method \\
Analysis type & Sensitivity: Compute required effect size \\ \hline
\multicolumn{2}{l}{\textit{Input Parameters}} \\
Tail(s) & Two \\
Parent distribution & Normal \\
$\alpha$ err prob & 0.005 \\
Power ($1-\beta$ err prob) & 0.99 \\
Sample size group 1 ($n_1$) & 11,918 \\
Sample size group 2 ($n_2$) & 253,445 \\ \hline
\multicolumn{2}{l}{\textit{Output Parameters}} \\
Noncentrality parameter $\delta$ & 5.1334215 \\
Critical $t$ & 2.8070584 \\
Df & 253401 \\
\textbf{Effect size $d$} & \textbf{0.0492377} \\ 
\end{tabular}
\end{table}

\begin{table}[h]
\centering
\caption{Sensitivity Analysis Parameters and Results (G*Power) on notebooks submitted before and after the deadline of the competitions}
\label{tab:gpower_sensitivity_wilcoxon_deadline}
\begin{tabular}{ll}
\textbf{Parameter} & \textbf{Value} \\ \hline
\multicolumn{2}{l}{\textit{Test Configuration}} \\
Test family & $t$ tests \\
Statistical test & Means: Wilcoxon-Mann-Whitney test (two groups) \\
Options & A.R.E. method \\
Analysis type & Sensitivity: Compute required effect size \\ \hline
\multicolumn{2}{l}{\textit{Input Parameters}} \\
Tail(s) & Two \\
Parent distribution & Normal \\
$\alpha$ err prob & 0.005 \\
Power ($1-\beta$ err prob) & 0.99 \\
Sample size group 1 ($n_1$) & 214,294 \\
Sample size group 2 ($n_2$) & 51,069 \\ \hline
\multicolumn{2}{l}{\textit{Output Parameters}} \\
Noncentrality parameter $\delta$ & 5.1334215 \\
Critical $t$ & 2.8070584 \\
Df & 253,401 \\
\textbf{Effect size $d$} & \textbf{0.0258677} \\ 
\end{tabular}
\end{table}

\begin{table}[h]
\centering
\caption{Sensitivity Analysis Parameters and Results (G*Power)}
\label{tab:gpower_sensitivity}
\begin{tabular}{ll}
\textbf{Parameter} & \textbf{Value} \\ \hline
\multicolumn{2}{l}{\textit{Test Configuration}} \\
Test family & Exact \\
Statistical test & Correlation: Bivariate normal model \\
Options & Large sample approximation (Fisher $Z$) \\
Analysis type & Sensitivity: Compute required effect size \\ \hline
\multicolumn{2}{l}{\textit{Input Parameters}} \\
Tail(s) & Two \\
Effect direction & $r \geq \rho$ \\
$\alpha$ err prob & 0.005 \\
Power ($1-\beta$ err prob) & 0.95 \\
Total sample size ($n$) & 265,363 \\
Correlation $\rho$ $H_0$ & 0 \\ \hline
\multicolumn{2}{l}{\textit{Output Parameters}} \\
Critical $z$ (Lower/Upper) & $\pm$ 2,5758293 \\
\textbf{Detectable correlation ($\rho$ $H_1$)} & \textbf{0.008193233} \\ 
\end{tabular}
\end{table}

\begin{table}[h]
\centering
\caption{Sensitivity Analysis Parameters and Results (G*Power) for group 2}
\label{tab:gpower_sensitivity_updated}
\begin{tabular}{ll}
\textbf{Parameter} & \textbf{Value} \\ \hline
\multicolumn{2}{l}{\textit{Test Configuration}} \\
Test family & Exact \\
Statistical test & Correlation: Bivariate normal model \\
Options & Large sample approximation (Fisher $Z$) \\
Analysis type & Sensitivity: Compute required effect size \\ \hline
\multicolumn{2}{l}{\textit{Input Parameters}} \\
Tail(s) & Two \\
Effect direction & $r \geq \rho$ \\
$\alpha$ err prob & 0.005 \\
Power ($1-\beta$ err prob) & 0.95 \\
Total sample size ($n$) & 45,443 \\
Correlation $\rho$ $H_0$ & 0 \\ \hline
\multicolumn{2}{l}{\textit{Output Parameters}} \\
Critical $z$ (Lower/Upper) & $\pm$ 2,8070338 \\
\textbf{Detectable correlation ($\rho$ $H_1$)} & \textbf{0.0208815} \\ 
\end{tabular}
\end{table}

\begin{table}[h]
\centering
\caption{Sensitivity Analysis Parameters and Results (G*Power) group 2 sample competition expert}
\label{tab:gpower_sensitivity_comp_expert}
\begin{tabular}{ll}
\textbf{Parameter} & \textbf{Value} \\ \hline
\multicolumn{2}{l}{\textit{Test Configuration}} \\
Test family & Exact \\
Statistical test & Correlation: Bivariate normal model \\
Options & Exact distribution \\
Analysis type & Sensitivity: Compute required effect size \\ \hline
\multicolumn{2}{l}{\textit{Input Parameters}} \\
Tail(s) & Two \\
Effect direction & $r \geq \rho$ \\
$\alpha$ err prob & 0.005 \\
Power ($1-\beta$ err prob) & 0.95 \\
Total sample size ($n$) & 3,620 \\
Correlation $\rho$ $H_0$ & 0 \\ \hline
\multicolumn{2}{l}{\textit{Output Parameters}} \\
Critical $r$ (Lower/Upper) & $\pm$ 0.0466452 \\
\textbf{Detectable correlation ($\rho$ $H_1$)} & \textbf{0.0738835} \\ 
\end{tabular}
\end{table}

\begin{table}[h]
\centering
\caption{Sensitivity Analysis Parameters and Results (G*Power) group 2 sample competition non-experts}
\label{tab:gpower_sensitivity_comp_non_expert}
\begin{tabular}{ll}
\textbf{Parameter} & \textbf{Value} \\ \hline
\multicolumn{2}{l}{\textit{Test Configuration}} \\
Test family & Exact \\
Statistical test & Correlation: Bivariate normal model \\
Options & Large sample approximation (Fisher $Z$) \\
Analysis type & Sensitivity: Compute required effect size \\ \hline
\multicolumn{2}{l}{\textit{Input Parameters}} \\
Tail(s) & Two \\
Effect direction & $r \geq \rho$ \\
$\alpha$ err prob & 0.005 \\
Power ($1-\beta$ err prob) & 0.95 \\
Total sample size ($n$) & 41,823 \\
Correlation $\rho$ $H_0$ & 0 \\ \hline
\multicolumn{2}{l}{\textit{Output Parameters}} \\
Critical $z$ (Lower/Upper) & $\pm$ 2.8070338 \\
\textbf{Detectable correlation ($\rho$ $H_1$)} & \textbf{0.0217662} \\ 
\end{tabular}
\end{table}

\begin{table}[h]
\centering
\caption{Sensitivity Analysis Parameters and Results (G*Power) group 2 sample code experts}
\label{tab:gpower_sensitivity_code_expert}
\begin{tabular}{ll}
\textbf{Parameter} & \textbf{Value} \\ \hline
\multicolumn{2}{l}{\textit{Test Configuration}} \\
Test family & Exact \\
Statistical test & Correlation: Bivariate normal model \\
Options & Exact distribution \\
Analysis type & Sensitivity: Compute required effect size \\ \hline
\multicolumn{2}{l}{\textit{Input Parameters}} \\
Tail(s) & Two \\
Effect direction & $r \geq \rho$ \\
$\alpha$ err prob & 0.005 \\
Power ($1-\beta$ err prob) & 0.95 \\
Total sample size ($n$) & 6,529 \\
Correlation $\rho$ $H_0$ & 0 \\ \hline
\multicolumn{2}{l}{\textit{Output Parameters}} \\
Critical $r$ (Lower/Upper) & $\pm$ 0.0347357 \\
\textbf{Detectable correlation ($\rho$ $H_1$)} & \textbf{0.0550509} \\ 
\end{tabular}
\end{table}

\begin{table}[h]
\centering
\caption{Sensitivity Analysis Parameters and Results (G*Power) group 2 sample code non-experts}
\label{tab:gpower_sensitivity_code_non_expert}
\begin{tabular}{ll}
\textbf{Parameter} & \textbf{Value} \\ \hline
\multicolumn{2}{l}{\textit{Test Configuration}} \\
Test family & Exact \\
Statistical test & Correlation: Bivariate normal model \\
Options & Large sample approximation (Fisher $Z$) \\
Analysis type & Sensitivity: Compute required effect size \\ \hline
\multicolumn{2}{l}{\textit{Input Parameters}} \\
Tail(s) & Two \\
Effect direction & $r \geq \rho$ \\
$\alpha$ err prob & 0.005 \\
Power ($1-\beta$ err prob) & 0.95 \\
Total sample size ($n$) & 38,914 \\
Correlation $\rho$ $H_0$ & 0 \\ \hline
\multicolumn{2}{l}{\textit{Output Parameters}} \\
Critical $z$ (Lower/Upper) & $\pm$ 2.8070338 \\
\textbf{Detectable correlation ($\rho$ $H_1$)} & \textbf{0.0225649} \\ 
\end{tabular}
\end{table}

\begin{table}[h]
\centering
\caption{Sensitivity Analysis Parameters and Results (G*Power)}
\label{tab:gpower_wilcoxon_comp_rq3}
\begin{tabular}{ll}
\textbf{Parameter} & \textbf{Value} \\ \hline
\multicolumn{2}{l}{\textit{Test Configuration}} \\
Test family & $t$ tests \\
Statistical test & Means: Wilcoxon-Mann-Whitney test (two groups) \\
Options & A.R.E. method \\
Analysis type & Sensitivity: Compute required effect size \\ \hline
\multicolumn{2}{l}{\textit{Input Parameters}} \\
Tail(s) & Two \\
Parent distribution & Normal \\
$\alpha$ err prob & 0.005 \\
Power ($1-\beta$ err prob) & 0.95 \\
Sample size group 1 ($n_1$) & 3,620 \\
Sample size group 2 ($n_2$) & 41,823 \\ \hline
\multicolumn{2}{l}{\textit{Output Parameters}} \\
Noncentrality parameter $\delta$ & 4.4520895 \\
Critical $t$ & 2.8071774 \\
Df & 43,392.87 \\
\textbf{Effect size $d$} & \textbf{0.0789314} \\ 
\end{tabular}
\end{table}

\begin{table}[h]
\centering
\caption{Sensitivity Analysis Parameters and Results (G*Power)}
\label{tab:gpower_wilcoxon_code}
\begin{tabular}{ll}
\textbf{Parameter} & \textbf{Value} \\ \hline
\multicolumn{2}{l}{\textit{Test Configuration}} \\
Test family & $t$ tests \\
Statistical test & Means: Wilcoxon-Mann-Whitney test (two groups) \\
Options & A.R.E. method \\
Analysis type & Sensitivity: Compute required effect size \\ \hline
\multicolumn{2}{l}{\textit{Input Parameters}} \\
Tail(s) & Two \\
Parent distribution & Normal \\
$\alpha$ err prob & 0.005 \\
Power ($1-\beta$ err prob) & 0.95 \\
Sample size group 1 ($n_1$) & 6,529 \\
Sample size group 2 ($n_2$) & 38,914 \\ \hline
\multicolumn{2}{l}{\textit{Output Parameters}} \\
Noncentrality parameter $\delta$ & 4.4520895 \\
Critical $t$ & 2.8071774 \\
Df & 43,392.87 \\
\textbf{Effect size $d$} & \textbf{0.0609306} \\ 
\end{tabular}
\end{table}

\section{Most violated Pylint rules in our dataset}
\label{sec:appendix:pylint_violations}

\begin{longtable}{c lcc}
\caption{Top 20 rules violated in the \textit{Full Dataset} ordered by mean density (mean expressed in $\times 10^{-3}$).}
\label{tab:pylint_top20} \\

\hline
\# & \textit{Rule} & \textit{Mean ($\times10^{-3}$)} & \textit{\%} \\
\hline
\endfirsthead

\multicolumn{4}{c}{\tablename\ \thetable\ -- continued from previous page} \\
\hline
\# & \textit{Rule} & \textit{Mean ($\times10^{-3}$)} & \textit{\%} \\
\hline
\endhead

\hline
\multicolumn{4}{r}{continued on next page} \\
\endfoot

\hline
\endlastfoot

\multicolumn{4}{l}{\textbf{Convention}} \\
\hline
1  & line-too-long (C0301)                  & 90.093 & 98.29 \\
2  & trailing-whitespace (C0303)             & 89.477 & 94.00 \\
3  & wrong-import-position (C0413)           & 48.532 & 74.66 \\
4  & invalid-name (C0103)                    & 23.202 & 73.25 \\
5  & missing-function-docstring (C0116)      & 19.790 & 77.61 \\
6  & wrong-import-order (C0411)              & 14.153 & 81.43 \\
7  & ungrouped-imports (C0412)               & 9.755  & 58.46 \\
8  & missing-module-docstring (C0114)        & 7.791  & 99.39 \\
9  & consider-using-f-string (C0209)         & 4.373  & 26.08 \\
10 & missing-class-docstring (C0115)         & 3.013  & 30.60 \\
11 & multiple-statements (C0321)             & 2.553  & 13.26 \\
12 & superfluous-parens (C0325)              & 1.580  & 15.89 \\
13 & consider-using-enumerate (C0200)        & 0.898  & 11.08 \\
14 & singleton-comparison (C0121)            & 0.734  & 6.94  \\
15 & multiple-imports (C0410)                & 0.606  & 9.24  \\
16 & import-outside-toplevel (C0415)         & 0.390  & 4.64  \\
17 & trailing-newlines (C0305)               & 0.261  & 3.30  \\
18 & consider-using-dict-items (C0206)       & 0.183  & 3.39  \\
19 & unnecessary-lambda-assignment (C3001)   & 0.149  & 2.06  \\
20 & unidiomatic-typecheck (C0123)           & 0.140  & 2.03  \\
\hline

\multicolumn{4}{l}{\textbf{Refactor}} \\
\hline
1  & too-few-public-methods (R0903)          & 1.625 & 21.10 \\
2  & too-many-locals (R0914)                 & 1.424 & 26.69 \\
3  & too-many-arguments (R0913)              & 1.144 & 20.70 \\
4  & too-many-positional-arguments (R0917)   & 1.138 & 20.64 \\
5  & consider-using-from-import (R0402)      & 1.102 & 16.79 \\
6  & no-else-return (R1705)                  & 0.844 & 14.77 \\
7  & use-dict-literal (R1735)                & 0.830 & 7.58  \\
8  & super-with-arguments (R1725)            & 0.665 & 10.21 \\
9  & unnecessary-comprehension (R1721)       & 0.553 & 7.33  \\
10 & consider-using-with (R1732)             & 0.505 & 5.19  \\
11 & too-many-instance-attributes (R0902)    & 0.321 & 8.97  \\
12 & use-list-literal (R1734)                & 0.280 & 2.41  \\
13 & too-many-statements (R0915)             & 0.247 & 8.20  \\
14 & too-many-branches (R0912)               & 0.185 & 6.35  \\
15 & consider-using-generator (R1728)        & 0.169 & 2.90  \\
16 & chained-comparison (R1716)              & 0.163 & 1.48  \\
17 & inconsistent-return-statements (R1710)  & 0.161 & 3.67  \\
18 & consider-using-in (R1714)               & 0.152 & 2.47  \\
19 & useless-object-inheritance (R0205)      & 0.078 & 1.87  \\
20 & too-many-nested-blocks (R1702)          & 0.067 & 1.70  \\
\hline

\multicolumn{4}{l}{\textbf{Warning}} \\
\hline
1  & unused-import (W0611)                   & 37.657 & 86.38 \\
2  & redefined-outer-name (W0621)            & 24.897 & 66.99 \\
3  & bad-indentation (W0311)                 & 10.049 & 11.45 \\
4  & reimported (W0404)                      & 9.193  & 43.34 \\
5  & unused-variable (W0612)                 & 2.510  & 29.22 \\
6  & unnecessary-semicolon (W0301)           & 2.271  & 10.69 \\
7  & unused-argument (W0613)                 & 2.048  & 21.58 \\
8  & f-string-without-interpolation (W1309)  & 1.122  & 9.98  \\
9  & pointless-string-statement (W0105)      & 0.980  & 7.40  \\
10 & wildcard-import (W0401)                 & 0.867  & 7.28  \\
11 & unspecified-encoding (W1514)            & 0.864  & 9.51  \\
12 & unnecessary-lambda (W0108)              & 0.753  & 6.46  \\
13 & attribute-defined-outside-init (W0201)  & 0.553  & 5.02  \\
14 & bare-except (W0702)                     & 0.505  & 7.92  \\
15 & anomalous-backslash-in-string (W1401)   & 0.459  & 2.76  \\
16 & redefined-builtin (W0622)               & 0.442  & 6.98  \\
17 & dangerous-default-value (W0102)         & 0.361  & 6.07  \\
18 & broad-exception-caught (W0718)          & 0.331  & 5.24  \\
19 & cell-var-from-loop (W0640)              & 0.309  & 3.15  \\
20 & undefined-loop-variable (W0631)         & 0.204  & 2.99  \\
\hline

\multicolumn{4}{l}{\textbf{Error}} \\
\hline
1  & undefined-variable (E0602)                       & 9.181 & 22.50 \\
2  & function-redefined (E0102)                       & 0.659 & 8.50  \\
3  & possibly-used-before-assignment (E0606)          & 0.385 & 7.57  \\
4  & used-before-assignment (E0601)                   & 0.133 & 2.70  \\
5  & too-many-function-args (E1121)                   & 0.098 & 1.14  \\
6  & invalid-character-zero-width-space (E2515)       & 0.085 & 0.43  \\
7  & unsubscriptable-object (E1136)                   & 0.077 & 0.90  \\
8  & no-self-argument (E0213)                         & 0.074 & 0.54  \\
9  & no-value-for-parameter (E1120)                   & 0.036 & 0.82  \\
10 & unsupported-binary-operation (E1131)             & 0.033 & 0.46  \\
11 & unsupported-assignment-operation (E1137)         & 0.029 & 0.32  \\
12 & access-member-before-definition (E0203)          & 0.020 & 0.17  \\
13 & bad-super-call (E1003)                           & 0.019 & 0.53  \\
14 & assignment-from-no-return (E1111)                & 0.012 & 0.25  \\
15 & not-callable (E1102)                             & 0.012 & 0.30  \\
16 & invalid-sequence-index (E1126)                   & 0.012 & 0.14  \\
17 & unexpected-keyword-arg (E1123)                   & 0.010 & 0.29  \\
18 & method-hidden (E0202)                            & 0.009 & 0.47  \\
19 & raising-bad-type (E0702)                         & 0.006 & 0.08  \\
20 & too-many-format-args (E1305)                     & 0.003 & 0.05  \\
\hline

\end{longtable}

\end{document}